\documentclass{article}
\usepackage[utf8]{inputenc}
\usepackage{import}
\usepackage{amsmath}
\usepackage{slashed}
\usepackage{amsfonts}
\usepackage{amssymb}
\usepackage{mathtools}
\usepackage{multirow}
\usepackage{caption}
\usepackage{authblk}
\usepackage{blindtext}
\usepackage{subcaption}
\usepackage{color}
\usepackage{setspace}

\DeclarePairedDelimiterX\braket[2]{\langle}{\rangle}{#1 \delimsize\vert #2}
\usepackage{abstract}
\usepackage{cite}

\usepackage{graphicx}
\title{\textbf{Light Cone Sum Rules and Form Factors for $p\rightarrow e^+ \gamma$}}
\author[1,2]{Anshika Bansal\thanks{ anshika@prl.res.in, anshika.bansal@iitgn.ac.in}}
\author[1]{Namit Mahajan\thanks{ nmahajan@prl.res.in}}
\affil[1]{Physical Research Laboratory, Ahmedabad, 380009, India.}
\affil[2]{Indian Institute of Technology, Gandhinagar, 382424, India.}
\date{}
\begin{document}
\maketitle
\doublespacing
\begin{abstract}
  Proton decay is a baryon number violating process, and hence is forbidden in the Standard Model (SM) of particle physics. Baryon number violation is expected to be an important criteria to explain the matter-anti-matter asymmetry of the universe. Any detection of the proton decay will be a direct evidence of physics beyond the SM. In SMEFT, proton decay is possible via baryon number violating dimension-6 operators. In this work, we pay attention to the decay channel $p\rightarrow e^+ \gamma$ which is expected to be an experimentally cleaner channel due to less nuclear absorption. The gauge invariant amplitude of this process involves two form factors. We calculate these form factors in the framework of light cone sum rules (LCSR) using photon DAs upto two particle twist-3 accuracy as well as proton DAs of twist-3 accuracy. We find that the form factors calculated using photon DAs are more reliable.
\end{abstract}
\section{Introduction}
\label{intro}
In particle physics, the Standard Model (SM) of strong and electro-weak interactions is the most successful model of particle interactions. In the SM, baryon number conservation is an accidental global symmetry at the classical level.
In 1967, Sakharov proposed that baryon number violation is one of the important criteria to explain the matter-anti matter asymmetry of the universe \cite{Sakharov:1967dj}. Baryon number violation at the perturbative level is well motivated in the theories of grand unification (GUTs), supersymmetry, models of baryogenesis, model building in string theory and in theories with extra dimensions, etc (see for example \cite{Pati:1973uk, Georgi:1974sy, Nilles:1983ge, Morrissey:2012db, FileviezPerez:2015mlm, Haber:1984rc, Chamoun:2020aft, Dienes:1998vh, Brandenberger:1988as, Fornal:2020esl, Murgui:2021bdy} and references therein). Proton decay is a baryon number violating process. Any observation of it is a direct indication of physics beyond the SM. 
This makes proton decay a crucial test of such models and an important window to understand the nature of matter unification. \\
In the case of GUTs, quarks and leptons fall in the common multiplets and hence can lead to proton decay at the tree level via the exchange of superheavy gauge bosons or scalar and/or vector leptoquarks. This makes it possible to write the effective baryon and lepton number violating operators of dim-6 by integrating out these heavy fields, such that they are consistent with the SM gauge symmetry. These effective operators are found to conserve $B-L$ which implies that a proton always decays into an antilepton (or antineutrino) (see \cite{Nath:2006ut,Langacker:1980js,Frampton:2021odk,Hisano:2022qll} for reviews on proton decays). \\
$p\rightarrow e^+ \pi^0$ is the most favoured channel in several GUTs models. As with any process involving hadrons, proton decay modes like $p\rightarrow e^+ \pi^0$ require hadronic matrix elements, the form factors, to be computed within some framework or at least properly estimated. This mode has been studied using various models of QCD, such as relativistic quark model, QCD sum rules, effective chiral theory, lattice QCD, \cite{Aoki:2006ib,Aoki:2017puj,Berezinsky:1981qb,Isgur:1982iz,Din:1979bz,Gavela:1981cf,Machacek:1979tx}. Very recently, it has been studied in the framework of light cone sum rules \cite{Haisch:2021hvj}. Another decay channel which is found to have strong constraints is the radiative mode: $p \rightarrow e^+ \gamma$. The radiative mode is expected to be suppressed by $\alpha_{em}$. In \cite{Silverman:1980ha}, it is been studied within SU(5) GUT set up. They pointed out that it might be a more feasible channel experimentally as there will be less nuclear absorption. The form factors have been evaluated with a simple harmonic oscillator potential as a model for binding the quarks inside the proton. In \cite{Eeg:1981be}, it was studied in the framework of bag model and they concluded that it is not a feasible channel for experiments as the decay rate is small. The experimental facilities have been advancing over the time (see \cite{Dev:2022jbf} for a review of different experiments and expected sensitivities expected at future experiments) and hence a reanalysis of this mode is required, including a fresh attempt at evaluation of the involved form factors.\\
Experimentally, Kolar Gold Field \cite{Krishnaswamy:1981uc}, NUSEX \cite{Battistoni:1985na}, SOUDAN \cite{Thron:1989cd}, Kamiokande \cite{Kamiokande-II:1989avz}, etc, were designed to detect the proton decay. At present, the Super-Kamiokande, the largest proton water Cherenkov detector, is the most sensitvite detector and has put the most stringent lower bounds on the partial life times for the proton decays, $\tau_p> 10^{34}$ \cite{Super-Kamiokande:2016exg}.  The lower bound for the radiative proton decay modes $p\rightarrow e^+ \gamma$ and $p\rightarrow \mu^+\gamma$ are $\tau_p>6.7\times 10^{32} \text{ and } \tau_p> 4.8\times 10^{32}$, respectively \cite{Zyla:2020zbs}. 
In the Water-Cherenkov experiments, such as Super-Kamiokande, the decay products of the proton are measured approximately at rest which makes the relevant energy scale for the process to be the proton mass (see \cite{Takhistov:2016eqm} for a review on Super-Kamiokande).\\ At these energy scales, a perturbative description for the hadronic transitions is not possible in QCD because of quark confinement. Hence, we need alternative ways to get an estimate of the hadronic matrix elements which can help us in probing the baryon-number violating physics with the help of experimental data. Light Cone Sum Rules (LCSR) is one such interesting framework which helps us to predict the hadronic matrix elements at the proton mass scales using the analytic properties of the correlation functions (see for example \cite{Colangelo:2000dp,Shifman:1978bx,Shifman:1998rb,Braun:1997kw,Khodjamirian:2020btr, narison1990qcd} for details). In this work, we study the $p\rightarrow e^+ \gamma$ in the framework of LCSR.\\
The rest of the paper is organised as follows: In Section-2, we discuss the general parametrisation of the amplitude for the decay in terms of the form factor and define the physical FFs involved. In Section-3, we discuss the computation of these form factors in the framework of LCSR. Here, we discuss the two cases: firstly, the use of photon distribution amplitudes and interpolating the proton state; and secondly, the use of proton distribution amplitudes and interpolating the photon state. In this section, we also discuss the numerical results obtained in both the cases. Section-4 is dedicated to discussion of the results and conclusions. This paper consists of five Appendices. In Appendix-\ref{DAs}, we collect the distribution amplitudes (DAs) of proton and photon upto the desired twist. Appendix-\ref{cf} and Appendix-\ref{cfpda} are dedicated to collect the correlation functions computed in QCD for the case employing photon DAs and proton DAs, respectively. In Appendix-\ref{appendixA}, we provide some useful identities and integrals along with definitions and conventions used through out the paper. Finally, we tabulate the numerical values of all the important parameters involved during numerical analysis in Appendix-\ref{appendixB}.

\section{Amplitude Computation}
\label{Amplitude}
Proton decay is a baryon number violating process. Though baryon number is a good symmetry in the SM, one can write higher dimensional effective operators which allow the proton to decay.  In a beyond the SM scenario, like GUTs, proton decay is possible even at tree level via an exchange of heavy gauge bosons or leptoquarks. On integrating out these heavy particles, one obtains the 
 baryon number violating dim-6 SMEFT lagrangian which preserves the $SU(3)_C\times SU(2)_L \times U(1)_Y$ invariance \cite{Weinberg:1979sa,Wilczek:1979hc,Claudson:1981gh,Chadha:1983sj}.
 \begin{equation}
    \mathcal{L}_{\slashed B}^{(6)}=\sum_{\Gamma,\Gamma'}c_{\Gamma\Gamma'}\mathcal{O}_{\Gamma\Gamma'}=\sum_{\Gamma,\Gamma'}c_{\Gamma\Gamma'}\epsilon^{abc}\left(\bar{d_a^c}P_\Gamma u_b\right)\left(\bar{e^c}P_{\Gamma'}u_c\right)
    \label{lag}
\end{equation}
Here, $\Gamma$,$\Gamma' \in \{L,R\}$ are the chirality projections. $c_{\Gamma\Gamma'}$ are the Wilson coefficients. $C=i \gamma^2\gamma^0$ is the charge conjugation matrix and $a,b,c$ are the colour indices. It is worth pointing out at this juncture that the above effective lagrangian is assumed to be expressed in terms of the physical quark and lepton fields at the relevant scale. This means that all the flavour mixing and perturbative renormalization group (RG) effects together with the short distance information, are collectively lumped in the Wilson coefficients $c_{\Gamma\Gamma'}$. Since the aim of the present work is to systematically evaluate  the corresponding form factors relevant for the radiative mode, the exact details of these effects are not particularly relevant here, and therefore not discussed further. It should be straightforward to explicitly express these dependencies in a concrete model of proton decay.\\ 
The transition amplitude for $p\rightarrow e^++\gamma$ is the matrix element of the dim-6 lagrangian given in Eqn-\ref{lag} between the initial and the final states. 
\begin{align}
    \mathcal{A}(p(p_p)\rightarrow e^+(p_e)\gamma(k)) &= \sum_{\Gamma \Gamma'} c_{\Gamma\Gamma'}\left<e^+(p_e)\gamma(k)\left|\mathcal{O}_{\Gamma\Gamma'}\right|p(p_p)\right> \nonumber \\ &= \sum_{\Gamma \Gamma'} c_{\Gamma\Gamma'}\left<e^+(p_e)\gamma(k)\left|\epsilon^{abc}\left(\bar{d_a^c}P_\Gamma u_b\right)\left(\bar{e^c}P_{\Gamma'}u_c\right)\right|p(p_p)\right>
\end{align}
As mentioned above, all the flavour effects are absorbed in the Wilson coefficients, $c_{\Gamma\Gamma'}$. 
On demanding the gauge invariance, this amplitude can be parametrised as
\begin{align}
    \mathcal{A}(p(p_p)\rightarrow e^+(p_e)\gamma(k)) &= \sum_{\Gamma\Gamma'} c_{\Gamma\Gamma'} \bar v_e^c P_{\Gamma'}\left\{i\sigma^{\alpha\beta} k_\beta\epsilon_{\alpha^*}A_{\Gamma\Gamma'}\right\} u_p(p_p).
    \label{ff}
\end{align}
where $A_{\Gamma\Gamma'}$ are the non-perturbative form factors. Parity conservation in QCD relates the different form factors relevant for the process:
\begin{eqnarray}
    A_{LL}&=-A_{RR} \hspace{2cm} A_{LR}=-A_{RL}.
\end{eqnarray}
Hence, this process involves only two independent gauge invariant form factors. For the present study, we choose them to be $A_{LL}$ and $A_{LR}$. Clearly, the main hurdle in obtaining the branching ratio is the knowledge of the form factors. All other factors are known once a given model of particle physics leading to proton decay is chosen. \\
The photon can be emitted either from the proton or the positron. The photon emission from positron can be trivially calculated and is not explicitly written as it does not contribute to the dipole transition depicted above. The photon emission from proton involves the photon emission from both u and d-quarks and contributes to the form factors. The study of these FFs in the framework of LCSR is the subject of the present study. The transition matrix element for the photon emission from proton can be factorised 
in the leptonic and hadronic parts as
 \begin{equation}
 \left<e^+(p_e)\gamma(k)\left|\mathcal{O}_{\Gamma\Gamma'}\right|p(p_p)\right>= \bar{v_e^c}(p_e) H_{\Gamma\Gamma'}(p_P,p_e)u_p(p_p).
\end{equation}
We choose to parametrise the hadronic matrix element $H_{\Gamma\Gamma'}u_p(p_p)$ as (see \cite{Deshpande:1981zq} for general parametrisation of the vertex for $b\to s\gamma$ transition): 
\begin{align}
     H_{\Gamma\Gamma'}(p_P,p_e)u_p(p_p)&= \left<\gamma(k)\left|\epsilon^{abc}\left(d_a^TCP_\Gamma u_b\right)\left(P_{\Gamma'}u_c\right)\right|p(p_p)\right>  \nonumber \\ &= P_{\Gamma'}\epsilon^*_\mu\left[F_{\Gamma\Gamma'}^1 \frac{\slashed k p_p^\mu}{m_p^2}+F_{\Gamma\Gamma'}^2 \frac{\slashed k k^\mu}{m_p^2}+F_{\Gamma\Gamma'}^3 \gamma^\mu +i F_{\Gamma\Gamma'}^4 \frac{\sigma^{\mu\nu}k_{\nu}}{m_p} +F_{\Gamma\Gamma'}^5 \frac{p_p^\mu}{m_p}+F_{\Gamma\Gamma'}^6 \frac{ k^\mu}{m_p}\right] u_p(p_p)
  \label{hadmax}
\end{align}
The physical FFs, $A_{\Gamma\Gamma'}$ are then related to $F_{\Gamma\Gamma'}^n$, with $n=\left\{1,2,3,4,5\right\}$, considering positron to be massless as,
\begin{equation}
\label{fphys}
    A_{\Gamma\Gamma'} = \frac{F^1_{\Gamma\Gamma'}}{2}+F^4_{\Gamma\Gamma'}+\frac{F^5_{\Gamma\Gamma'}}{2}.
\end{equation}

\section{Form Factors in the LCSR framework}
\label{FF}
To compute the FFs, $A_{\Gamma\Gamma'}$, in LCSR framework, we need to compute the hadronic matrix element given in Eq.(\ref{hadmax}) in QCD. For that there are two possibilities:
\begin{enumerate}
\item Interpolating the proton state and using the photon distribution amplitudes (DAs).
\item Interpolating the photon state and using the proton distribution amplitudes (DAs).
\end{enumerate}
We will discuss here both these approaches one by one with an aim to be finally able to compare the outcomes from both in order to gain deeper insights into the underlying non-perturbative dynamics.

\subsection{Case-1: Using proton interpolation and photon DAs}
\label{case1}
The interpolation current for the proton state is not unique. For the present study, we choose it to be
\begin{align}
    \chi(x)&= \epsilon^{abc}\left(u^{aT}(x)C \gamma_\mu u^b(x)\right)\gamma_5 \gamma^\mu d^c(x).
    \label{ioffe}
\end{align}
Here $C$ is the charge conjugation matrix, $\{a,b,c\}$ are the color indices and the superscript T denotes the transpose.
This current is popularly known as the Ioffe current \cite{Ioffe:1982ce} and is defined such that,
\begin{equation}
    \left<0\left|\chi(0)\right|p(p_p)\right>= m_p \lambda_p u_p(p_p)
\end{equation}
where, $m_p$ is the mass of proton, $u_p(p_p)$ is the proton spinor and $\lambda_p$ is the interaction strength of this interpolation current with the proton state.\\
In literature, this current is found to provide the maximum stability against the Borel mass, the parameter introduced in LCSR computations \cite{Braun:2006hz}. The Ioffe current is a linear combination of
\begin{equation}
    \chi_1(x)=\epsilon^{abc}\left(u^{Ta}(x)C\gamma_5 d^b(x)\right)u^c(x) \hspace{1cm } \text{and} \hspace{1cm}   \chi_2(x)=\epsilon^{abc}\left(u^{Ta}(x)C d^b(x)\right)\gamma_5 u^c(x)
    \label{chi12}
\end{equation}
such that $\chi(x) = 2(\chi_2-\chi_1)$ after performing Fierz transformation (see for example \cite{Leinweber:1994nm}). $\chi_1$ is the common choice of interpolation current employed in Lattice QCD computations.

On interpolating the proton state using the Ioffe current, the correlation function to be computed reads as
\begin{equation}
    \Pi_{\Gamma \Gamma'}(p_p,p_e) = i \int d^4x e^{i p_e.x} \left<\gamma(k)\left|T\{Q_{\Gamma\Gamma'}(x)\bar \chi(0)\}\right|0\right>.
    \label{corr}
\end{equation}
Here, $\bar \chi(0)\equiv \chi^\dagger(0)\gamma^0$, $Q_{\Gamma\Gamma'}(x) = \epsilon^{abc}\left(d_a^T C P_\Gamma u_b\right)\left(P_{\Gamma'} u_C\right)$ and $T$ denotes the time ordering.\\
One can get the  hadronic parametrisation of this correlation function by inserting a complete set of intermediate states with the same quantum numbers as the proton and isolating the pole contribution of the proton state as,
\begin{align}
      \Pi_{\Gamma \Gamma'}^{had}(p_p,p_e) &= -\frac{m_p \lambda_p}{p_p^2-m_p^2} H_{\Gamma\Gamma'}(p_e,p_p) (\slashed p_p + m_p)+\ldots \nonumber \\ &= \epsilon^*_\mu P_{\Gamma'}\left[\Pi_{\Gamma\Gamma'}^{had,PK}\frac{\slashed k p_p^\mu}{m_p^2}+ \Pi_{\Gamma\Gamma'}^{had,KK}\frac{\slashed k k^\mu}{m_p^2}+\Pi_{\Gamma\Gamma'}^{had,V}\gamma^\mu+\Pi_{\Gamma\Gamma'}^{had,T}\frac{i \sigma^{\mu\nu}k_\nu}{m_p}+\Pi_{\Gamma\Gamma'}^{had,P}\frac{ p_p^\mu}{m_p}\nonumber \right. \\ &+ \left.\Pi_{\Gamma\Gamma'}^{had,K}\frac{k^\mu}{m_p}+\Pi_{\Gamma\Gamma'}^{had,KPP}\frac{\slashed k p_p^\mu\slashed p_p}{m_p^3} + \Pi_{\Gamma\Gamma'}^{had,KKP}\frac{k^\mu \slashed k \slashed p_p}{m_p^3}+\Pi_{\Gamma\Gamma'}^{had,VP}\frac{\gamma^\mu \slashed p_p}{m_p}+\Pi_{\Gamma\Gamma'}^{had,TP}\frac{i\sigma^{\mu\nu}k_\nu \slashed  p_p}{m_p^2}\nonumber \right. \\ &+ \left.\Pi_{\Gamma\Gamma'}^{had,PP}\frac{\slashed p_p p_p^\mu}{m_p^2}+\Pi_{\Gamma\Gamma'}^{had,KP}\frac{ k^\mu \slashed p_p}{m_p^2}\right].
      \label{haddec}
\end{align}
The ellipsis above represent the heavy states i.e. excited states and continuum, contributions. The 12 Dirac structures in Eq.(\ref{haddec}) can be used to derive the form factors $A_{LL}$ and $A_{LR}$.\\ $\Pi_{\Gamma\Gamma'}^{had,r}$ with $r=\left\{PK,KK,V,T,P,K,KPP,KKP,VP,TP,PP,KP\right\}$ are the scalar functions of $p_p^2$ and $P_e^2=-p_e^2$ and can be parametrised in terms of spectral densities using the dispersion relation given by,
\begin{equation}
    \Pi_{\Gamma\Gamma'}^{had,r}(p_p^2,P_e^2) = \int_{0}^\infty ds \frac{\rho_{\Gamma\Gamma'}^{had,r}(s,P_e^2)}{s-p_p^2}.
\end{equation}
where, $\rho_{\Gamma\Gamma'}^{had,r}(s,P_e^2)$ are the spectral densities given by,
\begin{equation}
    \rho_{\Gamma\Gamma'}^{had,r}(s,P_e^2)= \frac{1}{\pi}\text{Im}\Pi_{\Gamma\Gamma'}^{had,r}(s+i\epsilon,P_e^2)
\end{equation}
These spectral densities can also be written by separating the pole contribution and the heavy states contributions as
\begin{equation}
     \rho_{\Gamma\Gamma'}^{had,r}(s,P_e^2)= \lambda_p m_p^2 \delta(s-m_p^2) F_{\Gamma\Gamma'}^r(s,P_e^2)+ \rho_{\Gamma\Gamma'}^{heavy,r}(s,P_e^2).
\end{equation}
where $F_{\Gamma\Gamma'}^r(s,P_e^2)$ can be related to $F_{\Gamma\Gamma'}^n(s,P_e^2)$ for $s=m_p^2$ i.e proton being onshell which is ensured by the delta function. These relations reads as,
\begin{align}
\label{relations}
  & F_{\Gamma\Gamma'}^{PK}(s,P_e^2)= F_{\Gamma\Gamma'}^{KPP}(s,P_e^2)= F_{\Gamma\Gamma'}^{1}(s,P_e^2), \hspace{2cm}  F_{\Gamma\Gamma'}^{KK}(s,P_e^2)= F_{\Gamma\Gamma'}^{KKP}(s,P_e^2)= F_{\Gamma\Gamma'}^{2}(s,P_e^2),\nonumber \\
  & F_{\Gamma\Gamma'}^{V}(s,P_e^2)= F_{\Gamma\Gamma'}^{VP}(s,P_e^2)= F_{\Gamma\Gamma'}^{3}(s,P_e^2), \hspace{2.2cm}  F_{\Gamma\Gamma'}^{T}(s,P_e^2)= F_{\Gamma\Gamma'}^{TP}(s,P_e^2)= F_{\Gamma\Gamma'}^{4}(s,P_e^2),\nonumber \\
   & F_{\Gamma\Gamma'}^{P}(s,P_e^2)= F_{\Gamma\Gamma'}^{PP}(s,P_e^2)= F_{\Gamma\Gamma'}^{5}(s,P_e^2), \hspace{2.2cm}  F_{\Gamma\Gamma'}^{K}(s,P_e^2)= F_{\Gamma\Gamma'}^{KP}(s,P_e^2)= F_{\Gamma\Gamma'}^{6}(s,P_e^2).
\end{align}
Using the assumptions of the quark-hadron duality, the spectral desities of the heavy states, $\rho_{\Gamma\Gamma'}^{heavy,r}(s,P_e^2)$, can be approximated to the spectral densities computed using the quantum chromodynamics (QCD) as,
\begin{equation}
    \int_{s_0}^\infty ds \frac{\rho_{\Gamma\Gamma'}^{heavy,r}(s,P_e^2)}{s-p_p^2} \approx \int_{s_0}^\infty ds \frac{\rho_{\Gamma\Gamma'}^{QCD,r}(s,P_e^2)}{s-p_p^2} = \int_{s_0}^\infty ds \frac{1}{\pi}\frac{\text{Im}(\Pi_{\Gamma\Gamma'}^{QCD,r}(s,P_e^2))}{s-p_p^2}
\end{equation}
with $s_0$ being the continuum threshold which is a free parameter and is expected to be chosen below or equal to the lightest excitation but well above the ground state. In the present case, the lightest excitation state is the Roper resonance with mass of 1.44 GeV. To compute the contribution of the spectral densities due to heavier states one needs to compute the correlation functions $\Pi_{\Gamma\Gamma'}^{r}(s,P_e^2)$ in QCD.\\
In QCD, the time ordered product in Eq.(\ref{corr}) can be computed by partially contracting the quark fields as,
\begin{align}
    T\left\{Q_{\Gamma\Gamma'}(x) \bar \chi(0)\right\}= -\frac{1}{2}\epsilon^{lmn}\epsilon^{ijk}P_{\Gamma'}&\left[\left(\bar u_l(0)\Gamma_Au_i(x)\right)\left\{\Gamma_A\gamma_\mu \tilde S_{jm}^{(u)}(x)P_\Gamma S_{nk}^{(d)}(x)\gamma^\mu \gamma_5 + S_{jm}^{(u)}(x)\gamma_\mu \tilde \Gamma_A P_\Gamma S_{nk}^{(d)}(x) \gamma^\mu \gamma_5\right\} \nonumber \right. \\&+ \left.\left(\bar d_l(0)\Gamma_A d_i(x)\right) \left\{S_{kn}^{(u)}(x)\gamma_\mu \tilde S_{jm}^{(u)}(x) P_\Gamma \Gamma_A \gamma^\mu \gamma_5\right\}\right].
    \label{contractions}
\end{align}

Here, we have employed the completeness relation given by,
\begin{equation}
    q(x)\bar q(0)= \frac{-1}{4}\left(\bar q (0) \Gamma_A q(x) \right)\Gamma^A
\end{equation}
with, $q=\{u,d\}$ and the chosen basis of gamma matrices is
\begin{equation}
    \Gamma_A = \left\{1, \gamma_5, \gamma^\rho, i\gamma_\rho \gamma_5, \frac{1}{\sqrt{2}}\sigma^{\rho\sigma}\right\}.
\end{equation}
Further, $\tilde \Gamma_A = C \Gamma_A^T C^{-1} = \eta_i \Gamma_A$
with $C= i \gamma^2\gamma^0$ and,
\begin{equation} \eta_i= \begin{cases} 1, & \ \Gamma_A=1,i\gamma_5, \gamma_\mu\gamma_5 \\ -1, & \Gamma_A= \gamma_\mu, \sigma_{\mu\nu} \end{cases} \end{equation}
$S^{ij}(x)$ is the quark propagator at the light like separations. In the massless limit, it is given by,
\begin{equation}
    S_{ij}(x) = \frac{i \slashed x}{2 \pi^2 x^4} \delta_{ij} -\frac{ \left<\bar q q\right>}{12}\delta_{ij}\left(1+\frac{m_0^2 x^2}{16}\right)+\ldots
\end{equation}
Here, $\left<\bar q q\right>$ is the quark condensate. Ellipses denote higher terms with one or more gluon exchanges which are not considered in this work. $m_0$ is associated with the mixed condensate as
\begin{equation}
    \left<\bar q g_s G.\sigma q\right>= m_0^2 \left<\bar q q\right>
\end{equation}
where $G.\sigma = G_{\mu\nu} \sigma^{\mu\nu}$. After performing the partial integrals, we are left with the matrix elements of two or more particle (quarks and gluons) operators which had been found to be written in terms of light cone distribution amplitudes (DAs) of photon of varying twist \cite{Ball:2002ps}. In the present work, we only consider the two particle DAs of twist-2 and twist-3 and leave a more detailed analysis of three-particle twist-3 and higher twist DAs (which are expected to be small) for future works.\\
The definitions of the DAs are collected in Appendix-\ref{DAs}. It is important to note here that at twist-2 there is only one DA, $\phi_\gamma(u,\mu)$ which appears in the matrix element of two quark operator with $\Gamma_A = \frac{1}{\sqrt{2}}\sigma^{\rho\sigma}$. At twist-3, there are 2 two-particle DAs which appears for $\Gamma_A = \{\gamma_\rho, i \gamma_\rho \gamma_5\}$ (for details look at Appendix-\ref{DAs}). 

On substituting the partial contractions of the time ordered product of quarks ,Eq.(\ref{contractions}) and the two-particle twist-2 and twist-3 DAs of the photon and summing up all the contributions, we get the analytic structure of the correlation function defined in Eq.(\ref{haddec}) in QCD as,
\begin{align}
      \Pi_{\Gamma \Gamma'}^{QCD}(p_p,p_e)  &= \epsilon^*_\mu P_{\Gamma'}\left[\Pi_{\Gamma\Gamma'}^{QCD,PK}\frac{\slashed k p_p^\mu}{m_p^2}+ \Pi_{\Gamma\Gamma'}^{QCD,KK}\frac{\slashed k k^\mu}{m_p^2}+\Pi_{\Gamma\Gamma'}^{QCD,V}\gamma^\mu+\Pi_{\Gamma\Gamma'}^{QCD,T}\frac{i \sigma^{\mu\nu}k_\nu}{m_p}+\Pi_{\Gamma\Gamma'}^{QCD,P}\frac{ p_p^\mu}{m_p}\nonumber \right. \\ &+ \left.\Pi_{\Gamma\Gamma'}^{QCD,K}\frac{k^\mu}{m_p}\Pi_{\Gamma\Gamma'}^{QCD,KPP}\frac{\slashed k p_p^\mu\slashed p_p}{m_p^3} + \Pi_{\Gamma\Gamma'}^{QCD,KKP}\frac{k^\mu \slashed k \slashed p_p}{m_p^3}+\Pi_{\Gamma\Gamma'}^{QCD,VP}\frac{\gamma^\mu \slashed p_p}{m_p}+\Pi_{\Gamma\Gamma'}^{QCD,TP}\frac{i\sigma^{\mu\nu}k_\nu \slashed  p_p}{m_p^2}\nonumber \right. \\ &+ \left.\Pi_{\Gamma\Gamma'}^{QCD,PP}\frac{\slashed p_p p_p^\mu}{m_p^2}+\Pi_{\Gamma\Gamma'}^{QCD,KP}\frac{ k^\mu \slashed p_p}{m_p^2}\right]
      \label{piqcd}
\end{align}
$\Pi_{\Gamma\Gamma'}^{QCD,r}$ are the scalar functions of $p_p^2$ and $P_e^2$. The analytic expressions for these functions are lengthy and hence are provided in Appendix-\ref{cf}. We also provide several useful identities and integrals in Appendix-\ref{appendixA}. According to the light cone sum rule matching condition,
\begin{equation}
    \Pi_{\Gamma\Gamma'}^{had,r}(p_p^2,P_e^2)= \Pi_{\Gamma\Gamma'}^{QCD,r}(p_p^2,P_e^2)
\end{equation}
Using the above relations, the final sum rule for $F_{\Gamma\Gamma'}^{r}$ reads as
\begin{equation}
    \lambda_p m_p^2 \frac{F_{\Gamma\Gamma'}^{r}(s,P_e^2)}{m_p^2-p_p^2} = \int_0^{s_0} ds \frac{1}{\pi}\frac{\text{Im}\Pi_{\Gamma\Gamma'}^{r,QCD}(s,P_e^2)}{s-p_p^2}
\end{equation}
 To suppress the effect of the heavy states, we perform the Borel transformation with respect to $p_p^2$. After Borel transformation the sum rule reads as (see Appendix-\ref{appendixA} for details),
 \begin{equation}
     F_{\Gamma\Gamma'}^r(s_0,P_e^2)= \frac{e^{\frac{m_p^2}{M^2}}}{\lambda_pm_p^2}\int_0^{s_0}ds e^{-\frac{s}{M^2}}\frac{1}{\pi}\text{Im}\Pi_{\Gamma\Gamma'}^{QCD,r}(s,P_e^2)
 \end{equation}
Here $M$ is the Borel mass and $s_0$ is the continuum threshold. These are the artefacts of the LCSR method, and have to be fixed such that the sum rule is saturated with the ground state and the heavy state contributions are properly suppressed. A typical rule of the thumb is to try and obtain at least $70\%$ contribution to the correlation function from the ground state itself. 
The details on these parameters is given in the next section.

 \subsubsection{Numerical Analysis}
 \label{na}
\begin{figure}[h]
   \begin{subfigure}{0.5\textwidth}
    \centering
    \includegraphics[width=0.95\linewidth]{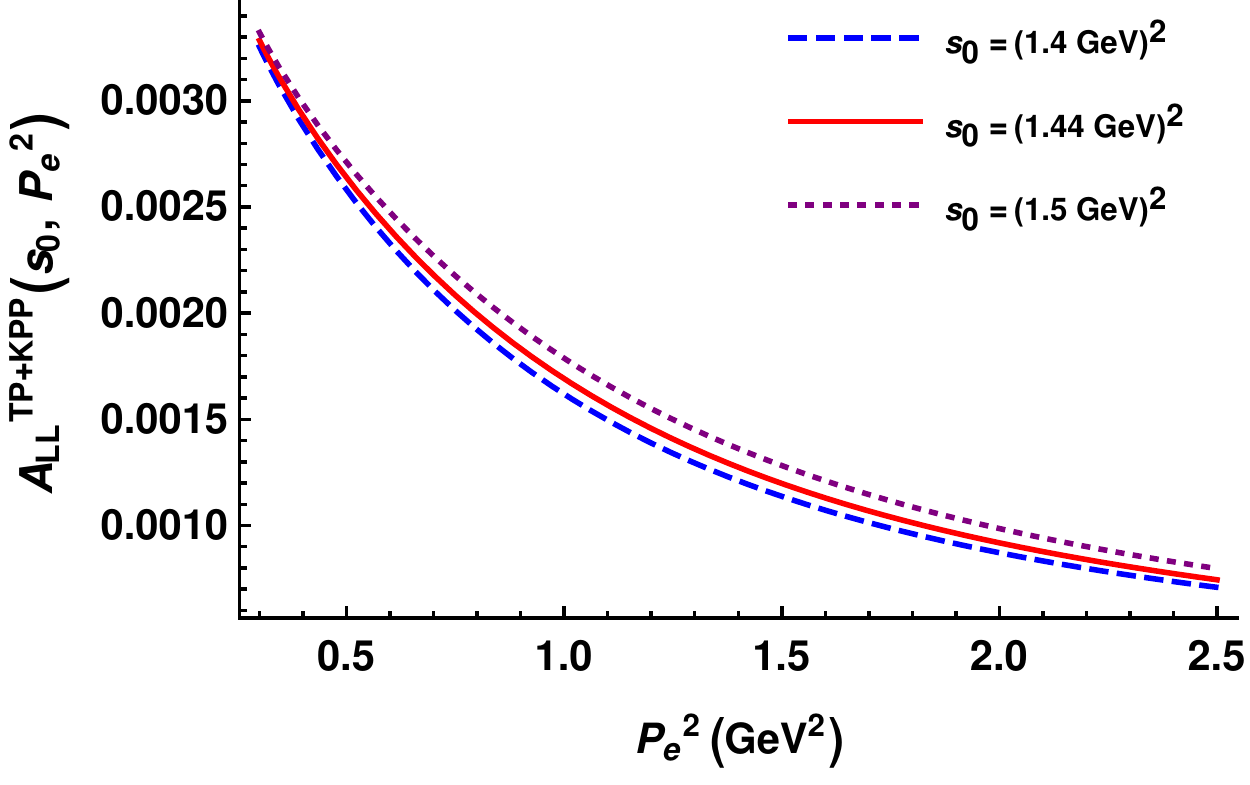}
    \caption{}
   \end{subfigure}
   \begin{subfigure}{0.5\textwidth}
    \centering
    \includegraphics[width=0.95\linewidth]{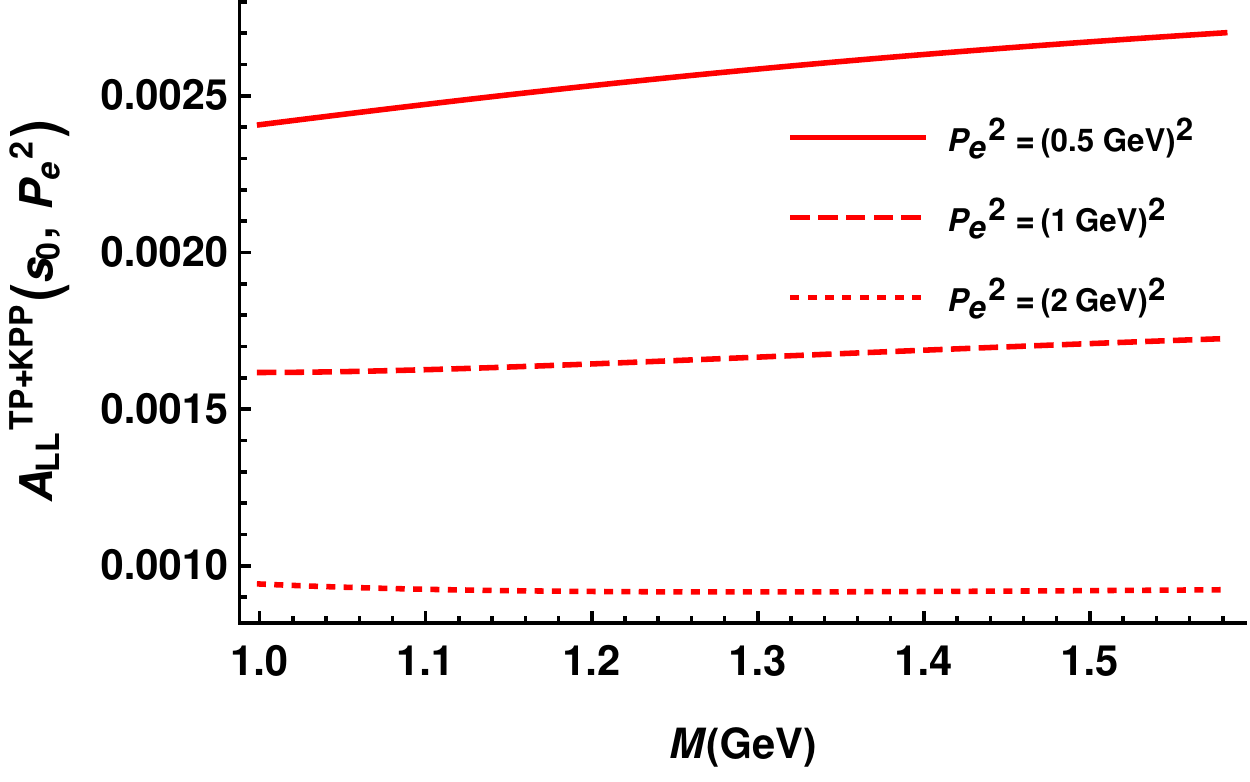}
    \caption{}
   \end{subfigure}
    \caption{The physical FF, $A_{LL}(s_0,P_e^2)$ is calculated from the combination of $F_{LL}^{TP}$ and $F_{LL}^{KPP}$ employing photon DAs. Left panel: $A_{LL}^{TP+KPP}(s_0,P_e^2)$ vs $P_e^2$ is shown for three values of $s_0= (1.4 \text{ GeV})^2$(violate dotted), $s_0= (1.44 \text{ GeV})^2$(red solid) and $s_0= (1.5 \text{ GeV})^2$ (blue dashed) at the Borel Mass, $M^2= 2 \text{ GeV}^2$. Right Panel: $A_{LL}^{TP+KPP}(s_0,P_e^2)$ vs $M$ is shown for three values of $P_e^2= 0.5 \text{ GeV}^2$(red solid), $P_e^2= 1 \text{ GeV}^2$(red dashed) and $P_e^2= 2 \text{ GeV}^2$ (red dotted) at the continuum threshold, $s_0=(1.44 \text{ GeV})^2$.}
    \label{LL}
\end{figure}

\begin{figure}[h]
   \begin{subfigure}{0.5\textwidth}
    \centering
    \includegraphics[width=0.95\linewidth]{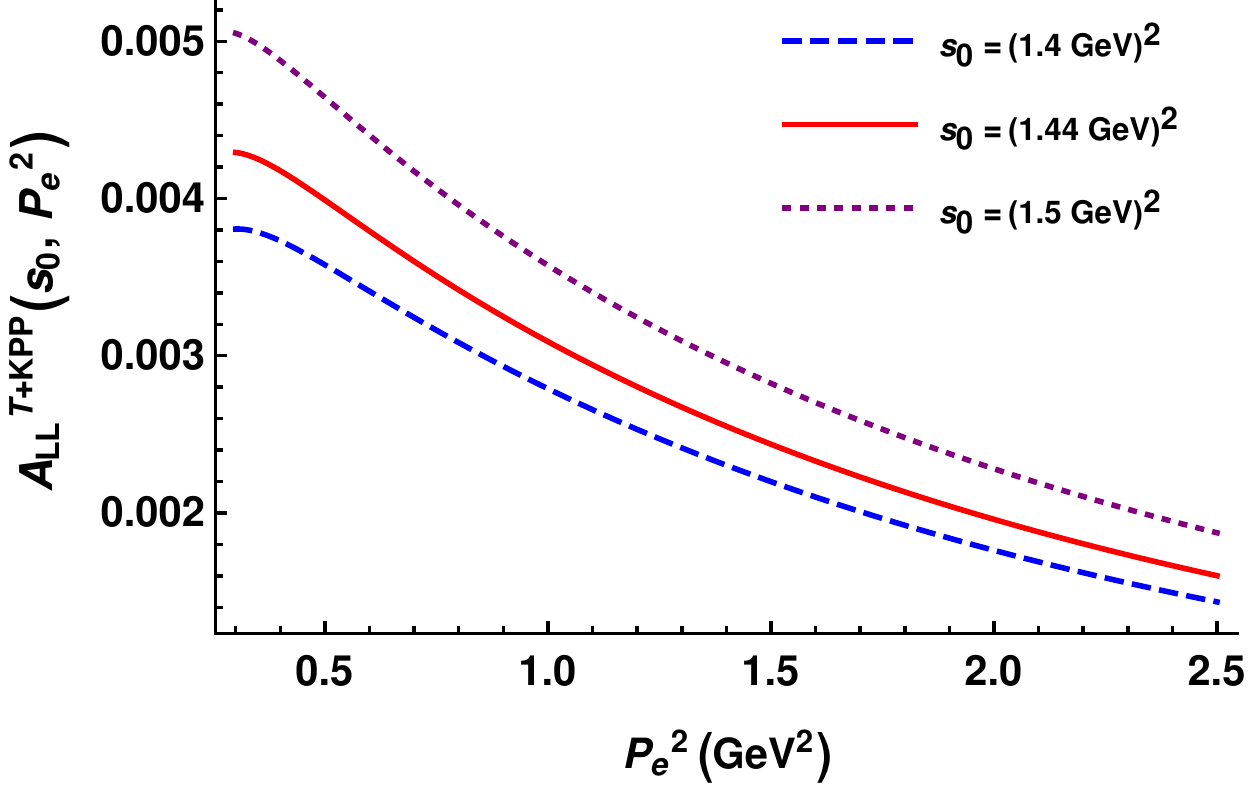}
    \caption{}
   \end{subfigure}
   \begin{subfigure}{0.5\textwidth}
    \centering
    \includegraphics[width=0.95\linewidth]{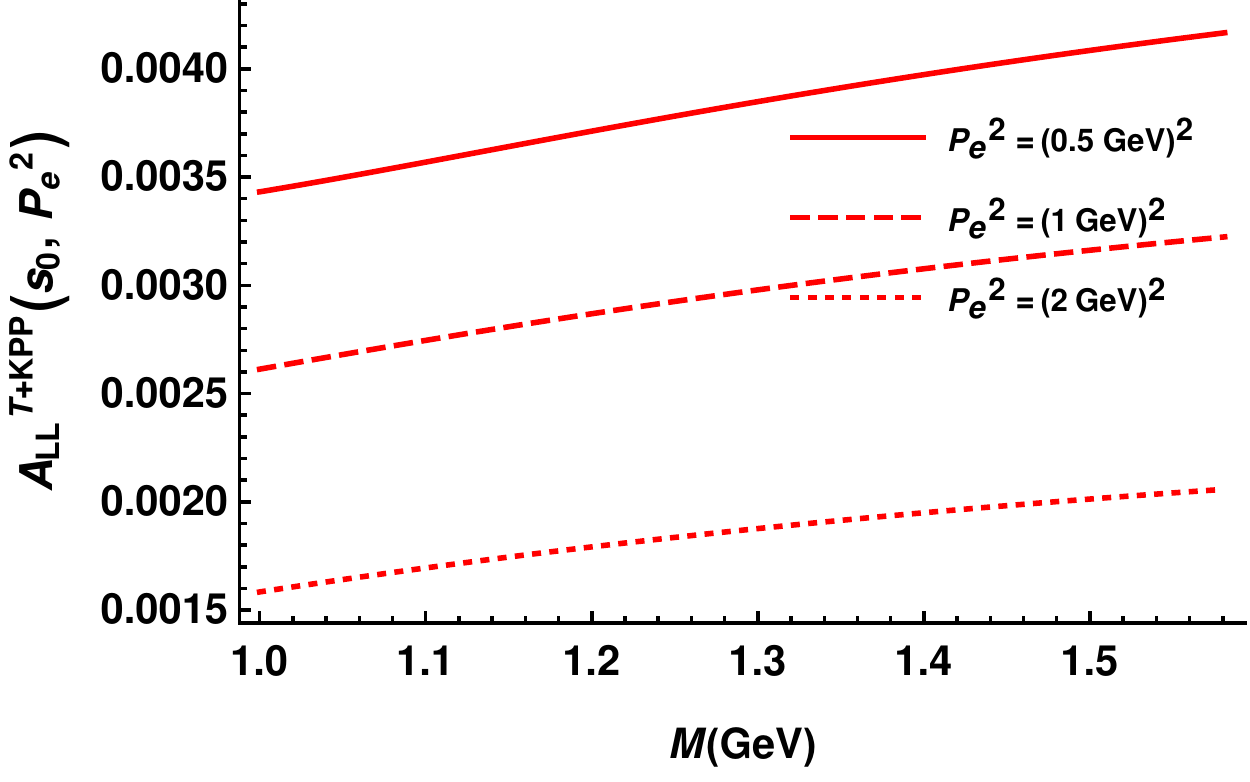}
    \caption{}
   \end{subfigure}
  \caption{Same as Fig(\ref{LL}) but now with the combinations of $F_{LL}^{T}$ and $F_{LL}^{KPP}$.}
    \label{LLTkpp}
\end{figure}

\begin{figure}
   \begin{subfigure}{0.5\textwidth}
    \centering
    \includegraphics[width=0.95\linewidth]{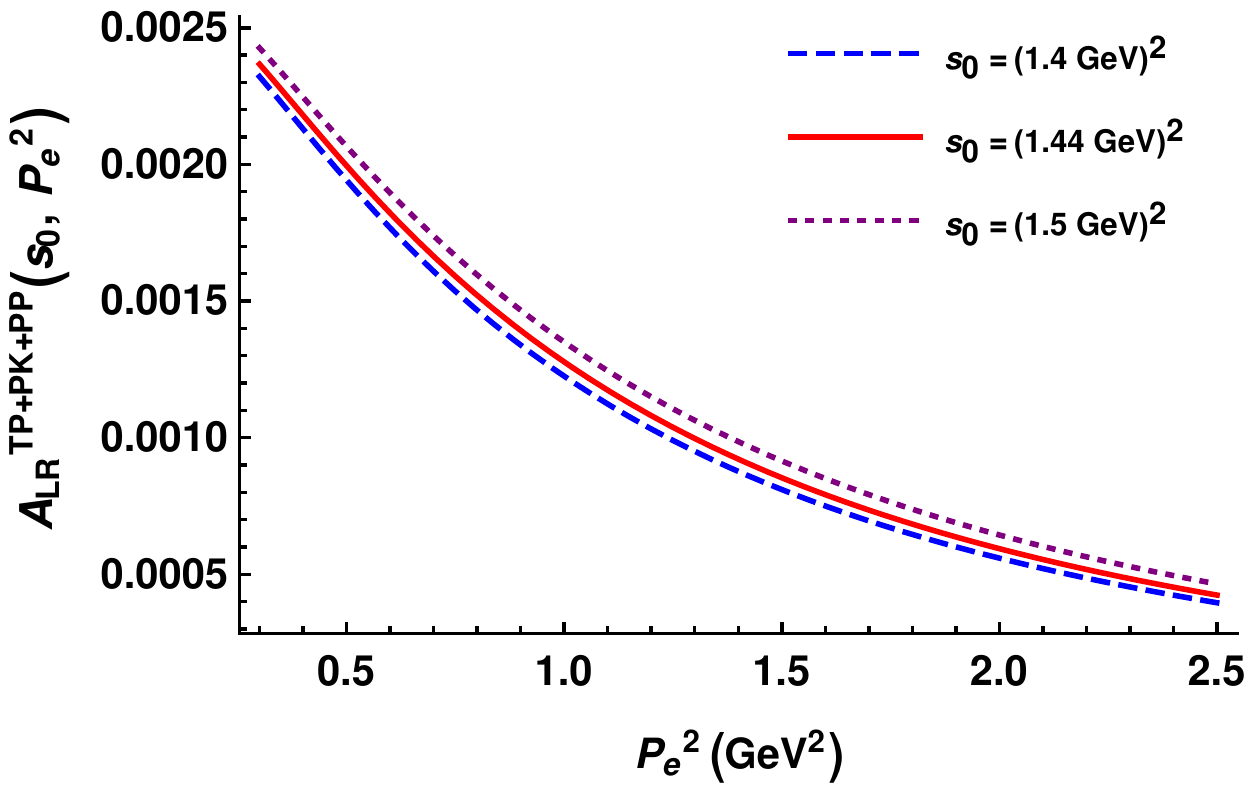}
    \caption{}
   \end{subfigure}
   \begin{subfigure}{0.5\textwidth}
    \centering
    \includegraphics[width=0.95\linewidth]{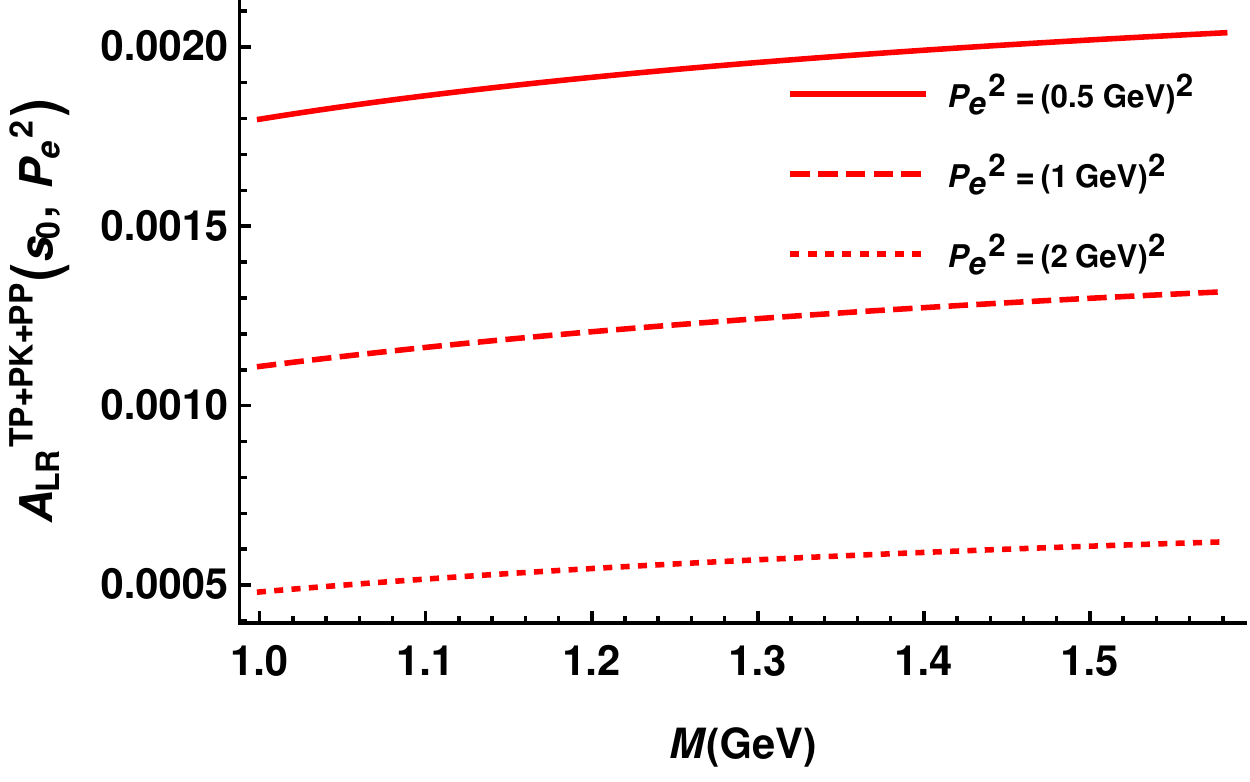}
    \caption{}
   \end{subfigure}
    \caption{The physical FF, $A_{LR}(s_0,P_e^2)$ is calculated from the combination of $F_{LR}^{TP}$, $F_{LR}^{PK}$ and $F_{LR}^{PP}$ employing photon DAs. Left panel: $A_{LR}^{TP+PK+PP}(s_0,P_e^2)$ vs $P_e^2$ is shown for three values of $s_0= (1.4 \text{ GeV})^2$(violate dotted), $s_0= (1.44 \text{ GeV})^2$(red solid) and $s_0= (1.5 \text{ GeV})^2$ (blue dashed) at the Borel Mass, $M^2= 2 \text{ GeV}^2$. Right Panel: $A_{LR}^{TP+PK+PP}(s_0,P_e^2)$ vs $M$ is shown for three values of $P_e^2= 0.5 \text{ GeV}^2$(red solid), $P_e^2= 1 \text{ GeV}^2$(red dashed) and $P_e^2= 2 \text{ GeV}^2$ (red dotted) at the continuum threshold, $s_0=(1.44 \text{ GeV})^2$.}
    \label{LRq21}
\end{figure}

\begin{figure}
   \begin{subfigure}{0.5\textwidth}
    \centering
    \includegraphics[width=0.95\linewidth]{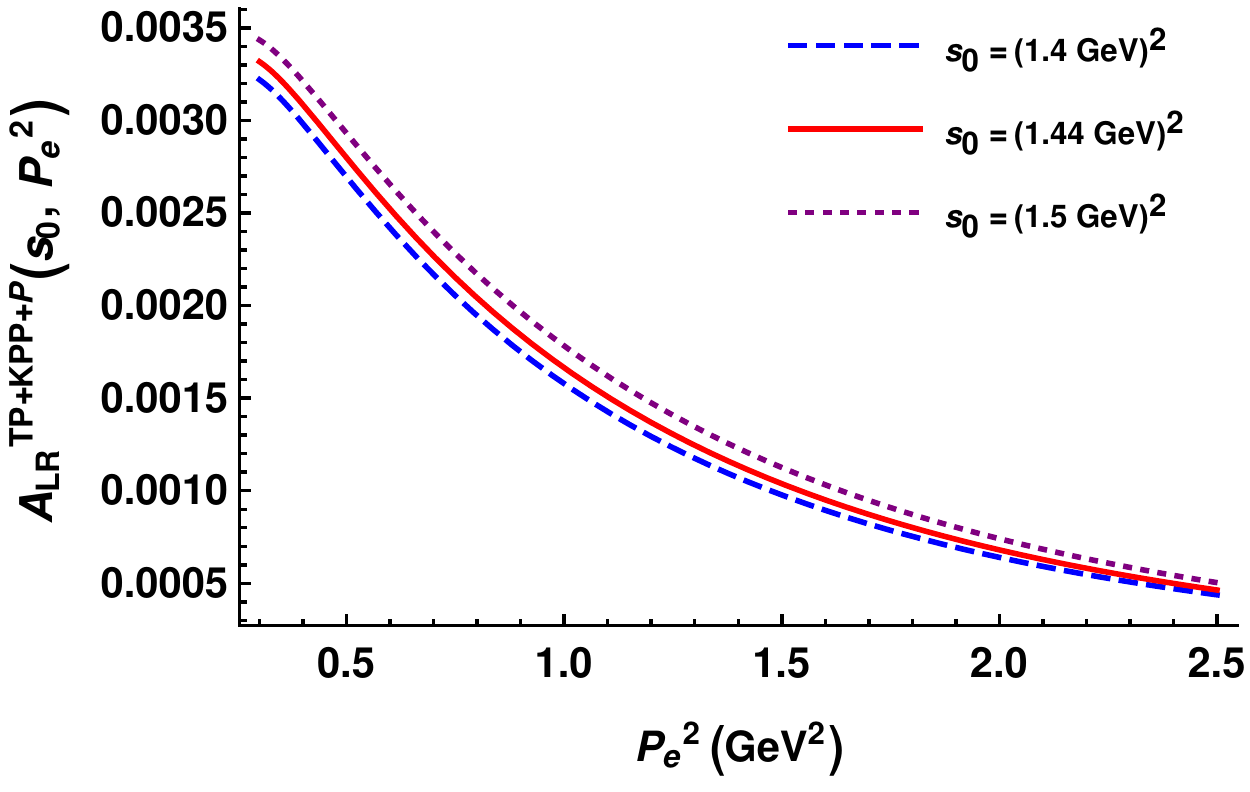}
    \caption{}
   \end{subfigure}
    \begin{subfigure}{0.5\textwidth}
    \centering
    \includegraphics[width=0.95\linewidth]{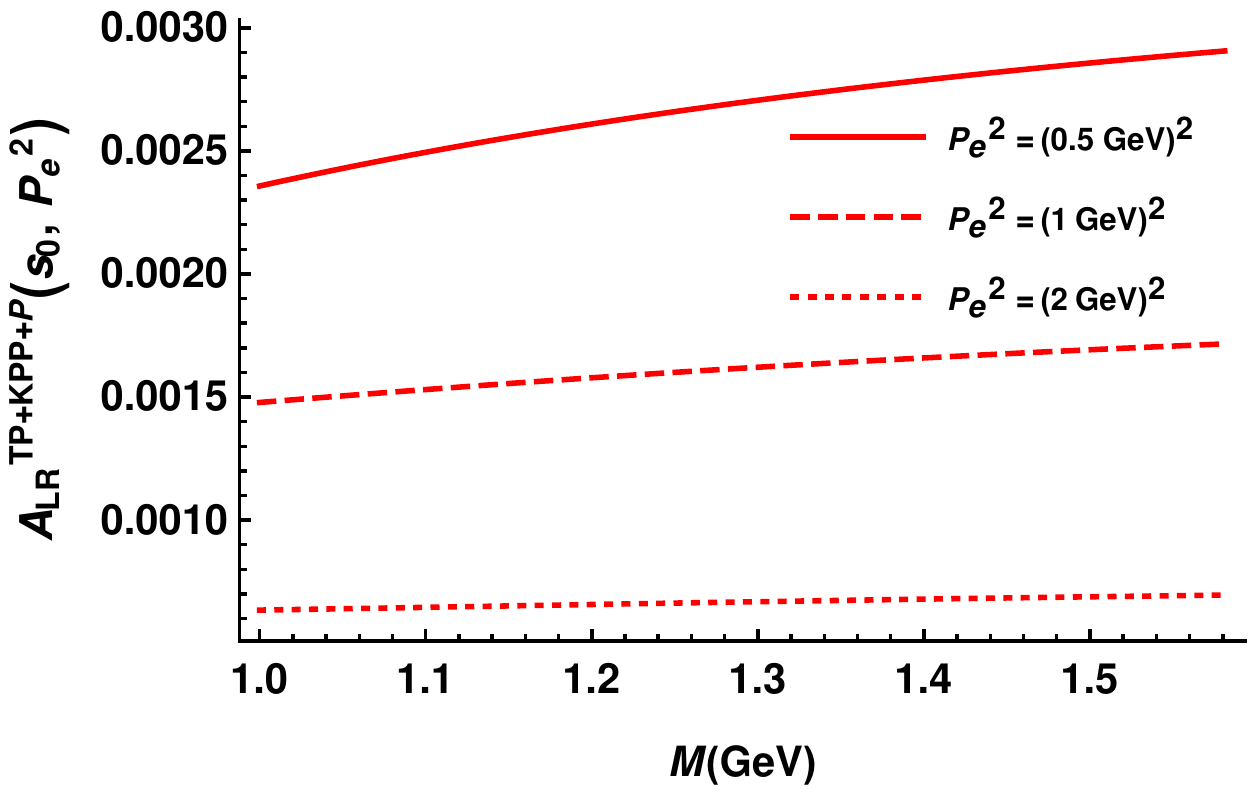}
    \caption{}
   \end{subfigure}
    \caption{Same as Fig(\ref{LRq21}) but now with the combinations of $F_{LR}^{TP}$, $F_{LR}^{KPP}$ and $F_{LR}^{P}$.}
    \label{LRq24}
\end{figure}

 The values of various parameters used during the numerical calculations are provided in Appendix-\ref{appendixB}. The physical FFs, $A_{\Gamma\Gamma'}$, for $\Gamma\Gamma'=LL$ and $LR$ are studied as a function of $P_e^2=-p_e^2$ and the Borel mass $M$. These FFs can be found from different combinations of $F_{\Gamma\Gamma'}$'s as can be read from Eq.(\ref{fphys}) and Eq.(\ref{relations}).  As the photon is onshell, we put $k^2=0$. For the case of $\Gamma\Gamma'=LL$, we have only two possibilities to extract $A_{LL}(s_0,P_e^2)$ which are from the combination of $F_{\Gamma\Gamma'}^T$ and $F_{\Gamma\Gamma'}^{TP}$ with $F_{\Gamma\Gamma'}^{KPP}$ as $F_{\Gamma\Gamma'}^{PK},F_{\Gamma\Gamma'}^{P}$, and $F_{\Gamma\Gamma'}^{PP}$ turns out to be zero in this case. In Fig.(\ref{LL}), we show the variation of $A_{LL}^{TP+KPP}(s_0,P_e^2)$ with $P_e^2$ for three different values of the continuum threshold $s_0$. In this Fig., we also show its variation with the Borel mass, $M$ for three different values of $P_e^2$ at fixed $s_0=(1.44 \text{GeV})^2$ which is equal to the Roper resonance. The combination  $A_{LL}^{T+KPP}(s_0,P_e^2)$ is found to be less stable when varying the parameters $s_0$ and $M$ (as can be seen from Fig.(\ref{LLTkpp}))  and hence is less reliable. On the face value, it is in broad agreement with $A_{LL}^{TP+KPP}$. As can be seen from the detailed expressions of these functions (listed in Appendix-\ref{cf}), condensate contributions are quite important (and also dominant in some cases), and therefore can't be simply ignored.
 For the case of $\Gamma\Gamma'=LR$, we have a total of eight combinations as can again be read from Eq.(\ref{fphys}) and Eq.(\ref{relations}). For this case as well, the four combinations which involves $F_{\Gamma\Gamma'}^T$ are found to be less stable against $s_0$ and $M$ and hence we do not show them here. The other four combinations involving $F_{\Gamma\Gamma'}^{TP}$ are shown in Fig.(\ref{LRq21})-Fig.(\ref{LRq24}).\\
\begin{figure}
   \begin{subfigure}{0.5\textwidth}
    \centering
    \includegraphics[width=0.95\linewidth]{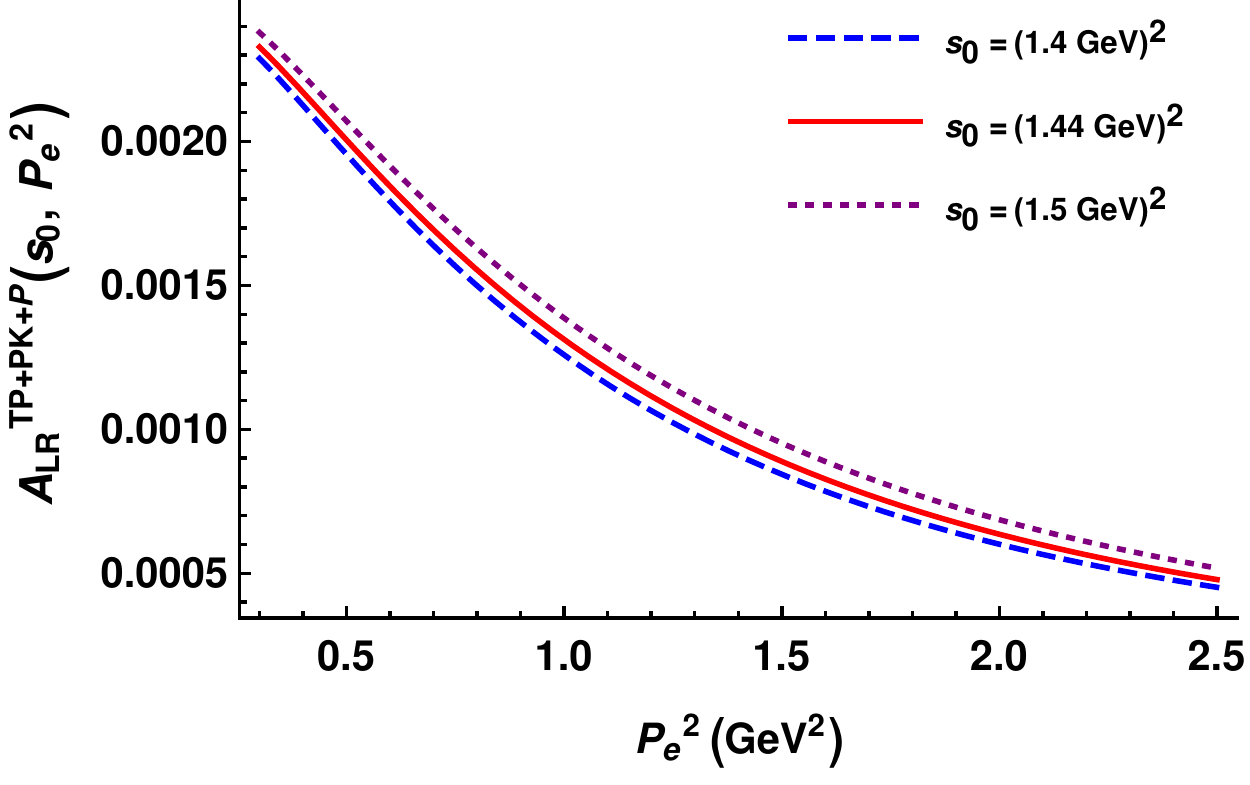}
    \caption{}
   \end{subfigure}
   \begin{subfigure}{0.5\textwidth}
    \centering
    \includegraphics[width=0.95\linewidth]{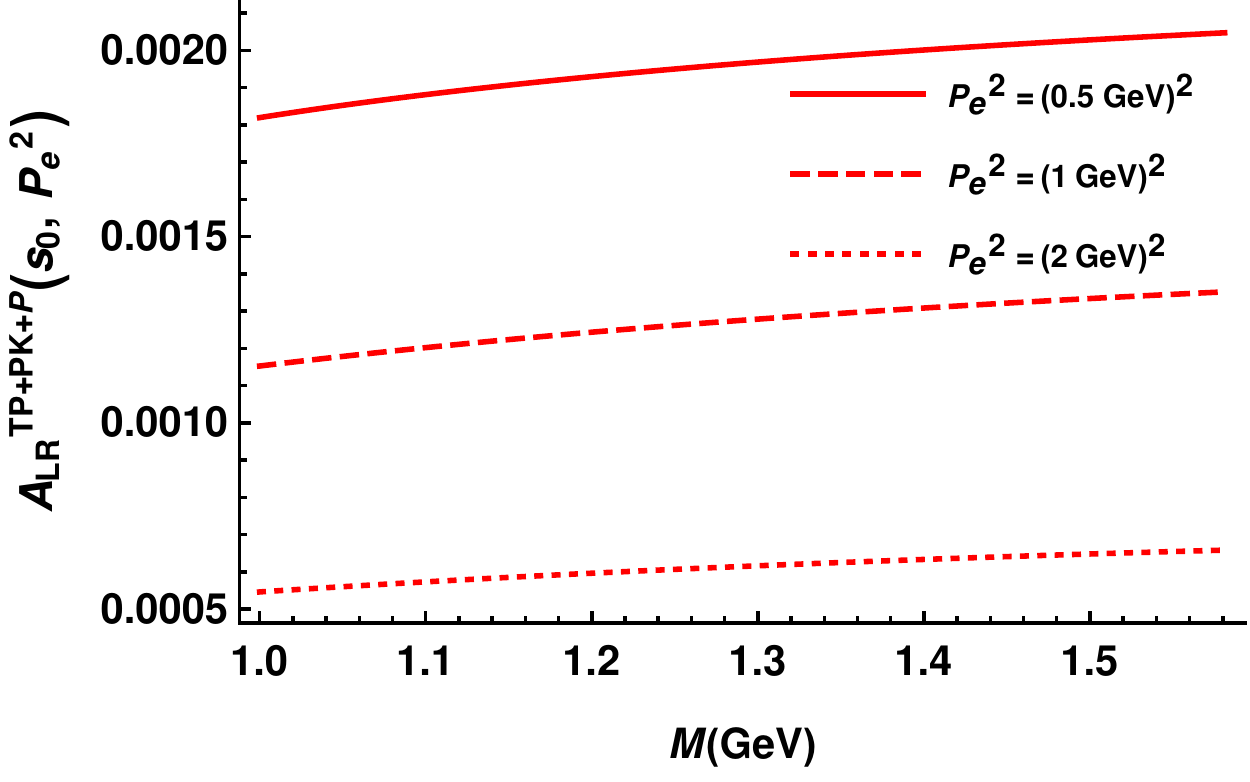}
    \caption{}
   \end{subfigure}
    \caption{Same as Fig(\ref{LRq21}) but now with the combinations of $F_{LR}^{TP}$, $F_{LR}^{PK}$ and $F_{LR}^{P}$.}
    \label{LRq22}
\end{figure}
\begin{figure}
   \begin{subfigure}{0.5\textwidth}
    \centering
    \includegraphics[width=0.95\linewidth]{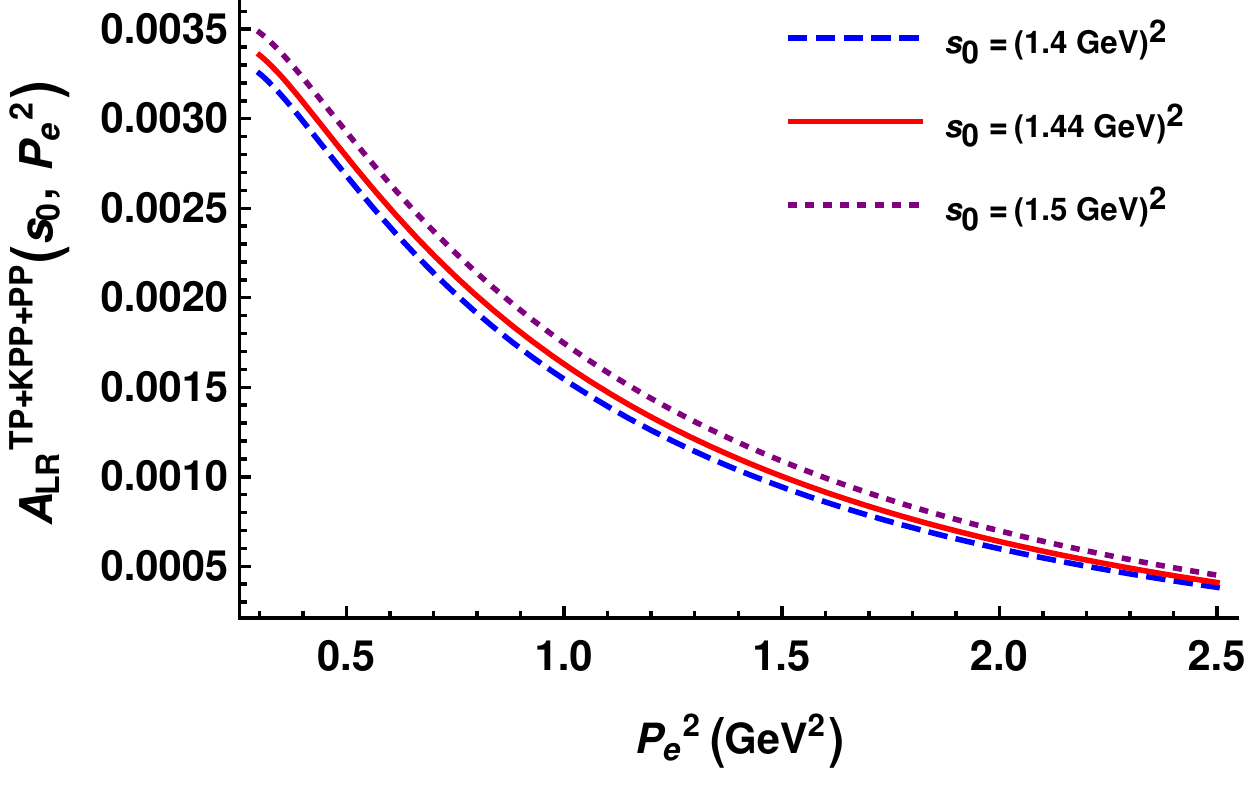}
    \caption{}
   \end{subfigure}
    \begin{subfigure}{0.5\textwidth}
    \centering
    \includegraphics[width=0.95\linewidth]{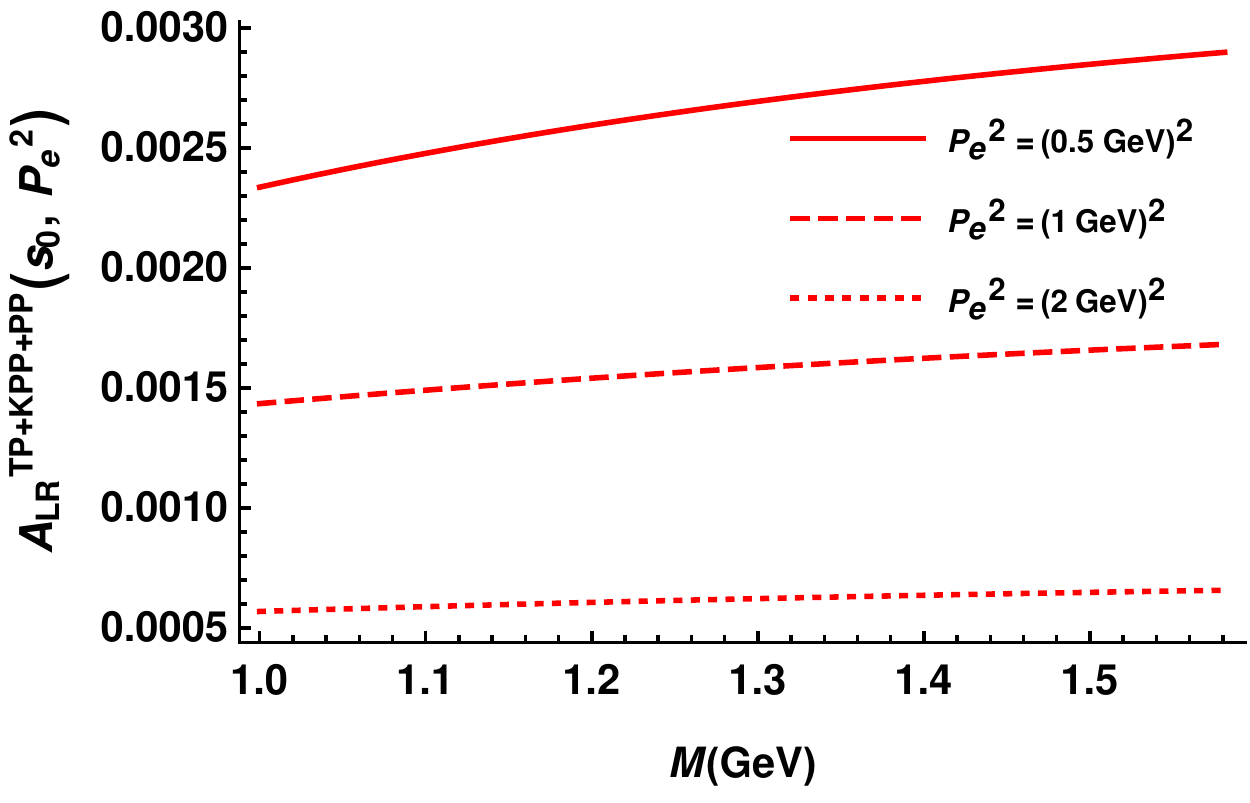}
    \caption{}
   \end{subfigure}
    \caption{Same as Fig(\ref{LRq21}) but now with the combinations of $F_{LR}^{TP}$, $F_{LR}^{KPP}$ and $F_{LR}^{PP}$.}
    \label{LRq23}
\end{figure}
 The values of the physical FFs, $A_{\Gamma\Gamma'}$ at $P_e^2=0.5 \text{ GeV}^2$\footnote{LCSR calculations are trustworthy at $\vert Q^2\vert\to\infty$, where $Q^2$ is the momentum transferred squared. To be consistent with this requirement, in this case, we have chosen $Q^2=P_e^2=0.5 \text{ GeV}^2$.} and $M^2=2 \text{ GeV}^2 $ for $s_0 \text{ }(=1.44 \text{ GeV})^2$ are found to be
 \begin{equation}
    A_{LL}^{T+KPP}(1.44^2,0.5) = 0.00399 \text{ GeV}^2, \hspace{1cm}  A_{LL}^{TP+KPP}(1.44^2,0.5)= 0.00264 \text{ GeV}^2
 \end{equation}
 \begin{equation}
     A_{LR}^{TP+KPP+P}(1.44^2,0.5) = 0.00280\text{ GeV}^2 , \hspace{1cm} A_{LR}^{TP+KPP+PP}(1.44^2,0.5) = 0.00279 \text{ GeV}^2
 \end{equation}
 \begin{equation}
     A_{LR}^{TP+PK+P}(1.4^2,0.5)= 0.00200 \text{ GeV}^2,  \hspace{1cm}  A_{LR}^{TP+PK+PP}(1.4^2,0.5) = 0.00199 \text{ GeV}^2
 \end{equation}
 From the above equations, it is clearly evident that there is quite good consistency in the form factor, $A_{LR}$, determined from different combinations.

\subsection{Case-2: Using photon interpolation and proton DAs}
\label{case2}
Having worked through the details with the proton state being interpolated, we next seek to determine the relevant form factors, but this time employing the distribution amplitudes of the proton. Then, on interpolating the photon state, the hadronic matrix element in Eq.(\ref{hadmax}) reads as, 
 \begin{equation}
    H_{\Gamma\Gamma'}(p_p,p_e)u_p(p_p) = -i e \epsilon^*_\alpha \int d^4 x e^{ik.x}\left<0\left|T\{j_{em}^\alpha(x) Q_{\Gamma\Gamma'}(0)\}\right|p(p_p)\right>
    \label{ginterpolate}
\end{equation}
where, $j_{em}^\alpha(x) = Q_d \bar d(x) \gamma^\alpha d(x) + Q_u \bar u(x) \gamma^\alpha u(x) - \bar e(x) \gamma^\alpha e(x)$ is the electromagnetic current and 
\begin{equation}
    Q_{\Gamma\Gamma' } =\epsilon^{abc}\left(d_a^TCP_\Gamma u_b\right)\left(P_{\Gamma'}u_c\right).
\end{equation}
 Using the generalized Fierz transformations \cite{Nieves:2003in}, it can be written as,
\begin{equation}
   Q_{LL}= \frac{\epsilon^{abc}}{4}\left(2(P_Ld_a)(\bar u_c^cP_Lu_b)-(\sigma_{\mu\nu}P_Ld_a)(\bar u_c^c\sigma_{\mu\nu}P_Lu_b)\right), \text{ and}
\end{equation}
\begin{equation}
   Q_{LR}= \frac{\epsilon^{abc}}{4}\left(2(\gamma_\mu P_Ld_a)(\bar u_c^c\gamma^\mu P_Lu_b)\right).
\end{equation}
As discussed above, to get the sum rule, we need to calculate the correlation function in Eq.(\ref{ginterpolate}) in QCD. To get the time ordered product of the 
electromagnetic current with $Q_{LL}$ and $Q_{RL}$ we need
\begin{align}
    T\{j_{em}^\alpha(x)  \left(\Gamma_A P_L d_a\right)\left(\bar u_c^c \Gamma^A P_L u_b\right)\} &=  \left[Q_u \left\{\left(C\gamma^\alpha \tilde S_{ic}^u(x)\Gamma_A P_L\right)^{BF} \left(\Gamma^A P_L\right)^{CD}\left(\left(u_i^T(x)\right)^Bu_b^F(0)d_a^D(0)\right) \right.\right. \nonumber \\ &+ \left.\left.\left(C\Gamma_A P_L S_{bi}(x)\gamma^\alpha\right)^{EB} \left(\Gamma^A P_L\right)^{CD} \left(\left(u_c^T(0)\right)^E u_b^B(x)d_a^D(0)\right)\right\} \right. \nonumber \\ &- \left.  Q_d \left\{\left(\Gamma_A S_{ai}^d(x)\gamma^\alpha\right)^{CB}\left(C\Gamma^AP_L\right)^{EF}\left(\left(u_c^{T}(0)\right)^{E}u_b^F(0)d_i^B(x)\right)\right\}\right] 
\end{align}
Here, capital alphabets ($E,F,B,C,D$) are the Dirac indices, $\{a,b,c,i\}$ are the color indices and superscript $T$ denotes the transpose. $\Gamma_A=\{1,\sigma_{\mu\nu}\}$ and $\Gamma_A=\{\gamma_\mu\}$ for the case of $LL$ and $LR$, respectively. 
The matrix element of the remaining three quark operator between the proton state and the vacuum can be parametrised in terms of proton DAs of varying twists \cite{Braun:2000kw}. In the present work, we consider only the leading twist-3 DAs ( given in Appendix-\ref{DAs}), which can be defined by,
\begin{equation}
   4 \left<0\left|\epsilon^{abc} u_\alpha^a(a_1x)u_\beta^b(a_2x)d_\gamma^c(a_3x)\right|P(p)\right> = \sum_i \mathcal{F}^i(\{a_1,a_2,a_3\},(p.x)) X_{\alpha\beta}^i Y_\gamma^i
   \label{pda}
\end{equation}
where,
\begin{center}
    \begin{tabular}{c|c|c}
    $\mathcal{F}^i$& $X_{\alpha\beta}$&$Y_\gamma$ \\
    \hline
     $\mathcal{V}_1$& $(\slashed{p_p}C)_{\alpha\beta}$&$(\gamma_5 u_p)_\gamma
     $\\
      $\mathcal{A}_1$& $(\slashed{p_p}\gamma_5C)_{\alpha\beta}$&$( u_p)_\gamma
     $\\
      $\mathcal{T}_1$& $(p_p^\nu i \sigma_{\mu\nu C})_{\alpha\beta}$&$(\gamma^\mu\gamma_5 u_p)_\gamma
     $\\
     \hline
\end{tabular}
\end{center}
such that 
\begin{equation}
     X_i^T= \begin{cases} X_i, & \mathcal{F}_i \in \mathcal{V}_i,\mathcal{T}_i \\ -X_i, & \mathcal{F}_i \in \mathcal{A}_i\end{cases} 
\end{equation}
where, superscript $T$ represents transpose. The DAs, $\mathcal{F}_i$, have the following symmetry under the exchange of $a_1$ and $a_2$,
\begin{equation}
    \mathcal{F}_i(\left\{a_1,a_2,a_3\right\},(p_p.x))= \begin{cases} \mathcal{F}_i(\left\{a_2,a_1,a_3\right\},(p_p.x)), & \mathcal{F}_i \in \mathcal{V}_i,\mathcal{T}_i \\ -\mathcal{F}_i(\left\{a_2,a_1,a_3\right\},(p_p.x)), & \mathcal{F}_i \in \mathcal{A}_i\end{cases} 
\end{equation}
and
\begin{equation}
    \mathcal{F}^i (\{a_1,a_2,a_3\},(p.x)) = \int_0^1 \mathcal{D}\alpha_i e^{-i \alpha_i a_ip.x}F^i(\alpha_1,\alpha_2,\alpha_3)
\end{equation}
with $\mathcal{D}\alpha_i = d\alpha_1 d\alpha_2 d\alpha_3 \delta(1-\alpha_1-\alpha_2-\alpha_3)$. \\
Using these DAs and considering the photon emission from the u- and d-quark, the correlation function in Eq.(\ref{ginterpolate}) turns out to be,
\begin{equation}
    H_{\Gamma\Gamma'}^{QCD}u_p(p_p)= \epsilon^*_\alpha P_{\Gamma'} \left[F_{\Gamma\Gamma'}^{1,QCD}\frac{p_p^\alpha \slashed k}{m_p^2}+F_{\Gamma\Gamma'}^{2,QCD}\frac{k^\alpha \slashed k}{m_p^2}+F_{\Gamma\Gamma'}^{3,QCD}\gamma^\alpha+F_{\Gamma\Gamma'}^{4,QCD}\frac{i \sigma^{\alpha\beta}k_\beta}{m_p}+F_{\Gamma\Gamma'}^{5,QCD}\frac{p_p^\alpha}{m_p}+F_{\Gamma\Gamma'}^{6,QCD}\frac{k^\alpha}{m_p}\right]
\end{equation}
Here, $F_{\Gamma\Gamma'}$ are the scalar functions of $P'^2=(p_p-k)^2$ and $K^2=-k^2$, and are provided in Appendix-\ref{cfpda}. 
Upon saturating with the intermediate lowest state, the hadronic decomposition reads as,
\begin{align}
    H^{had}_{\Gamma\Gamma'}u_p(p_p)&= -e \epsilon^*_\alpha  \frac{P_\Gamma'}{4} \lambda m_p \frac{\slashed p_p - \slashed k+m_p}{(p_p-k)^2-m_p^2}\left\{\gamma^\alpha W_1(K^2)-\frac{i\sigma^{\alpha\beta}k_\beta}{2m_p}W_2(K^2)\right\}u_p(p_p)+ \ldots \nonumber \\
    &= \epsilon^*_\alpha P_{\Gamma'} \left[F_{\Gamma\Gamma'}^{1,had}\frac{p_p^\alpha \slashed k}{m_p^2}+F_{\Gamma\Gamma'}^{2,had}\frac{k^\alpha \slashed k}{m_p^2}+F_{\Gamma\Gamma'}^{3,had}\gamma^\alpha+F_{\Gamma\Gamma'}^{4,had}\frac{i \sigma^{\alpha\beta}k_\beta}{m_p}+F_{\Gamma\Gamma'}^{5,had}\frac{p_p^\alpha}{m_p}+F_{\Gamma\Gamma'}^{6,had}\frac{k^\alpha}{m_p}\right]
\end{align}
Here, ellipses represent the contribution from the heavy states and $\lambda$ is the coupling strength of the proton interpolation current with the proton state. $\lambda=\lambda_p'$ and $\lambda=-\lambda_p$ for $\Gamma\Gamma' = LL$ and $\Gamma\Gamma'=LR$, respectively and are defined in Eq.(\ref{chi}) and Eq.(\ref{chip}), respectively. $W_1(K^2)$ and $W_2(K^2)$ are the electromagnetic electric and magnetic form factors of the proton and are defined as,
\begin{equation}
    \left<p(p_p-k)\left|j_\alpha^{em}(0)\right|p(p_p)\right>= \bar u_p(p_p-k)\left[W_1(K^2)\gamma_\alpha-i\frac{\sigma_{\alpha\beta}k^\beta}{2m_p}W_2(K^2)\right]u_p(p_p). 
\end{equation} 
The scalar functions $F_{\Gamma\Gamma'}^{had,n}$ (for $n=1,2,3,4,5,6$) of $P'^2$ and $K^2$ are related to $W_1(K^2) $ and $W_2(K^2)$ via following relations:
\begin{align}
    &F_{LL}^{1,had}= \frac{-e}{4}m_p^2 \lambda_p'\frac{W_2(K^2)}{P'^2-m_p^2} \hspace{4cm} F_{LL}^{2,had}= \frac{e}{4}m_p^2 \lambda_p'\frac{W_2(K^2)}{2\left(P'^2-m_p^2\right)} \hspace{2cm} \nonumber\\
   & F_{LL}^{3,had}= -\frac{e}{8} \lambda_p'W_2(K^2)\hspace{4.6cm} F_{LL}^{4,had}= \frac{e}{4}m_p^2 \lambda_p'\frac{W_1(K^2)+W_2(K^2)}{P'^2-m_p^2} \hspace{2cm} \nonumber\\ \hspace{2cm} 
    & F_{LL}^{5,had}= \frac{-e}{2}m_p^2 \lambda_p'\frac{W_1(K^2)}{P'^2-m_p^2}\hspace{4cm} F_{LL}^{6,had}= \frac{e}{4}m_p^2 \lambda_p'\frac{W_1(K^2)}{P'^2-m_p^2} \hspace{2cm}
    \label{relations2}
\end{align}
There will be similar relations between $F_{LR}^{n,had}$ and $W_{1,2}(K^2)$ with $\lambda_p'$ replaced by $-\lambda_p$. From Eq.(\ref{fphys}), we know $F_{\Gamma\Gamma'}^{1,4,5}$ are required to calculate the physical FFs, $A_{\Gamma\Gamma'}$. For $F_{\Gamma\Gamma'}^{1,4,5}$, after using the quark hadron duality and Borel transformation, the sum rule condition reads as,
\begin{align}
    F_{\Gamma\Gamma'}^{1,4,5}(s_0,K^2)= -\frac{\text{ Exp}\left(\frac{m_p^2}{M^2}\right)}{P'^2-m_p^2}\int_0^{s_0}ds \text{Exp}\left(\frac{-s}{M^2}\right)\frac{1}{\pi}\text{Im}\left(F_{\Gamma\Gamma'}^{\{1,4,5\},QCD}(s,K^2)\right)
\end{align}
\subsubsection{Numerical Analysis}
\begin{figure}[h]
   \begin{subfigure}{0.5\textwidth}
    \centering
    \includegraphics[width=0.95\linewidth]{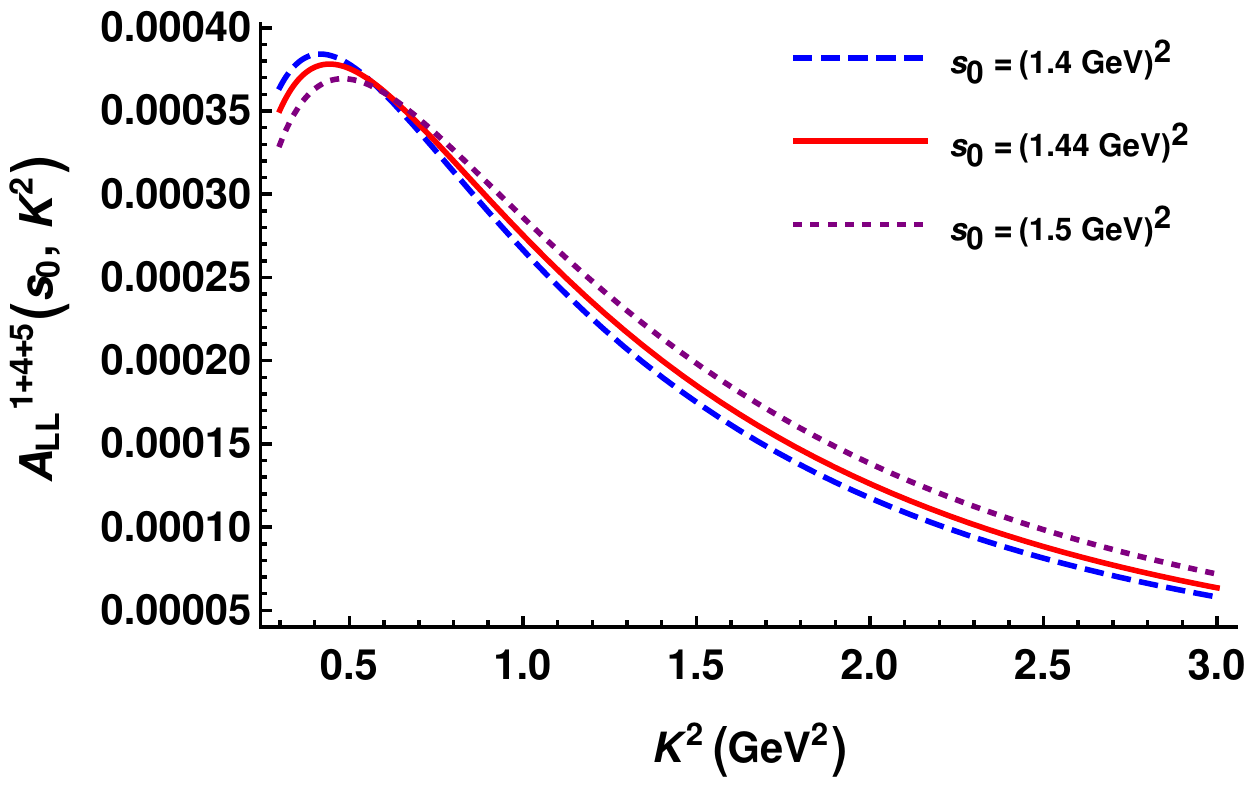}
    \caption{}
   \end{subfigure}
   \begin{subfigure}{0.5\textwidth}
    \centering
    \includegraphics[width=0.95\linewidth]{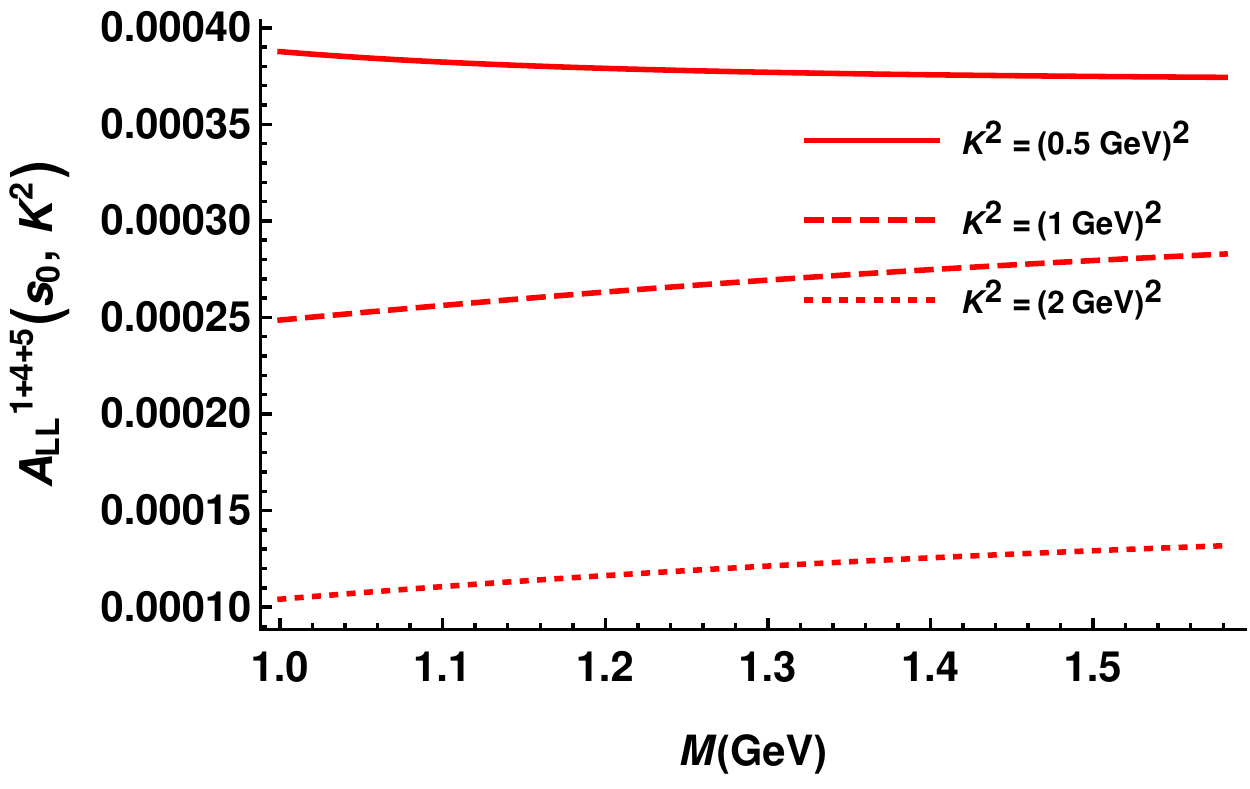}
    \caption{}
   \end{subfigure}
    \caption{The physical FF, $A_{LL}(s_0,K^2)$ is calculated from the combination of $F_{LL}^{1}$, $F_{LL}^{4}$ and $F_{LL}^{5}$ employing proton DAs. Left panel: $A_{LL}^{1+4+5}(s_0,K^2)$ vs $K^2$ is shown for three values of $s_0= (1.4 \text{ GeV})^2$(violate dotted), $s_0= (1.44 \text{ GeV})^2$(red solid) and $s_0= (1.5 \text{ GeV})^2$ (blue dashed) at the Borel Mass, $M^2= 2 \text{ GeV}^2$. Right Panel: $A_{LL}^{1+4+5}(s_0,K^2)$ vs $M$ is shown for three values of $K^2= 0.5 \text{ GeV}^2$(red solid), $K^2= 1 \text{ GeV}^2$(red dashed) and $K^2= 2 \text{ GeV}^2$ (red dotted) at the continuum threshold, $s_0=(1.44 \text{ GeV})^2$.}
    \label{LLp}
\end{figure}

\begin{figure}[h]
   \begin{subfigure}{0.5\textwidth}
    \centering
    \includegraphics[width=0.95\linewidth]{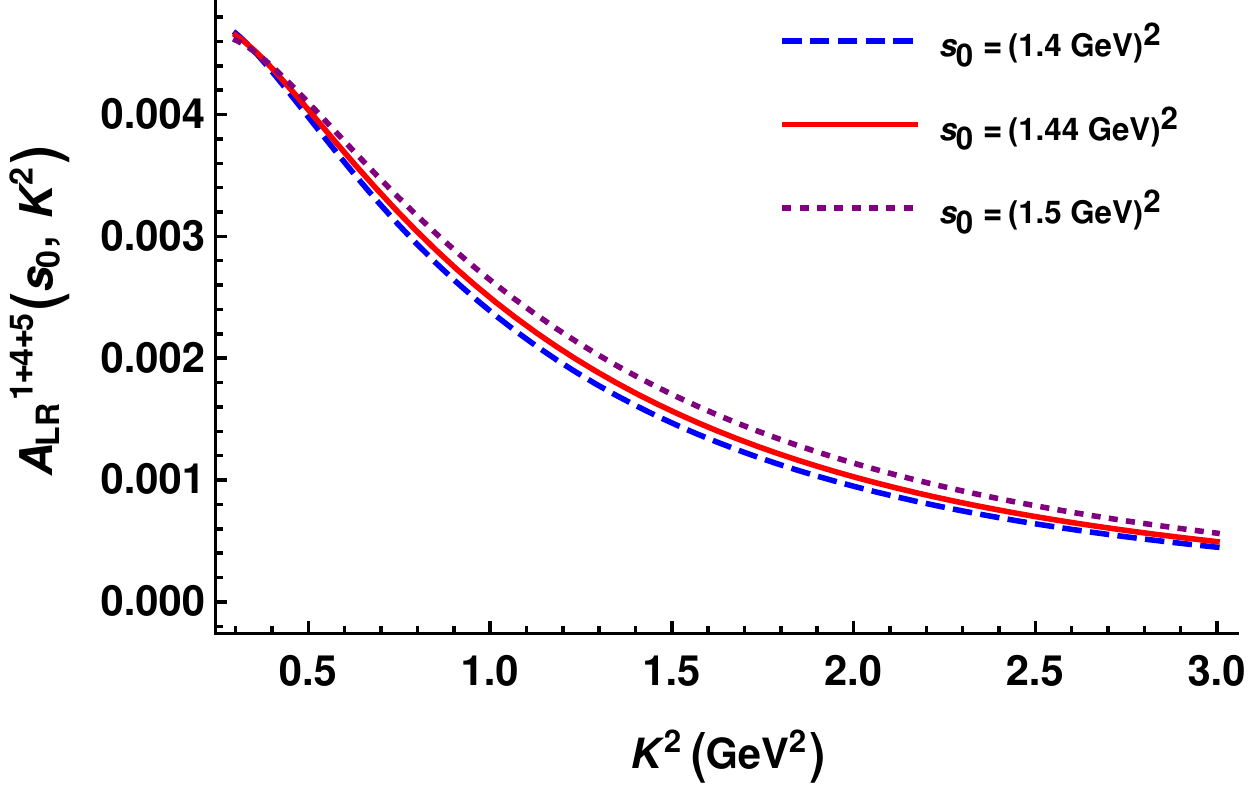}
    \caption{}
   \end{subfigure}
   \begin{subfigure}{0.5\textwidth}
    \centering
    \includegraphics[width=0.95\linewidth]{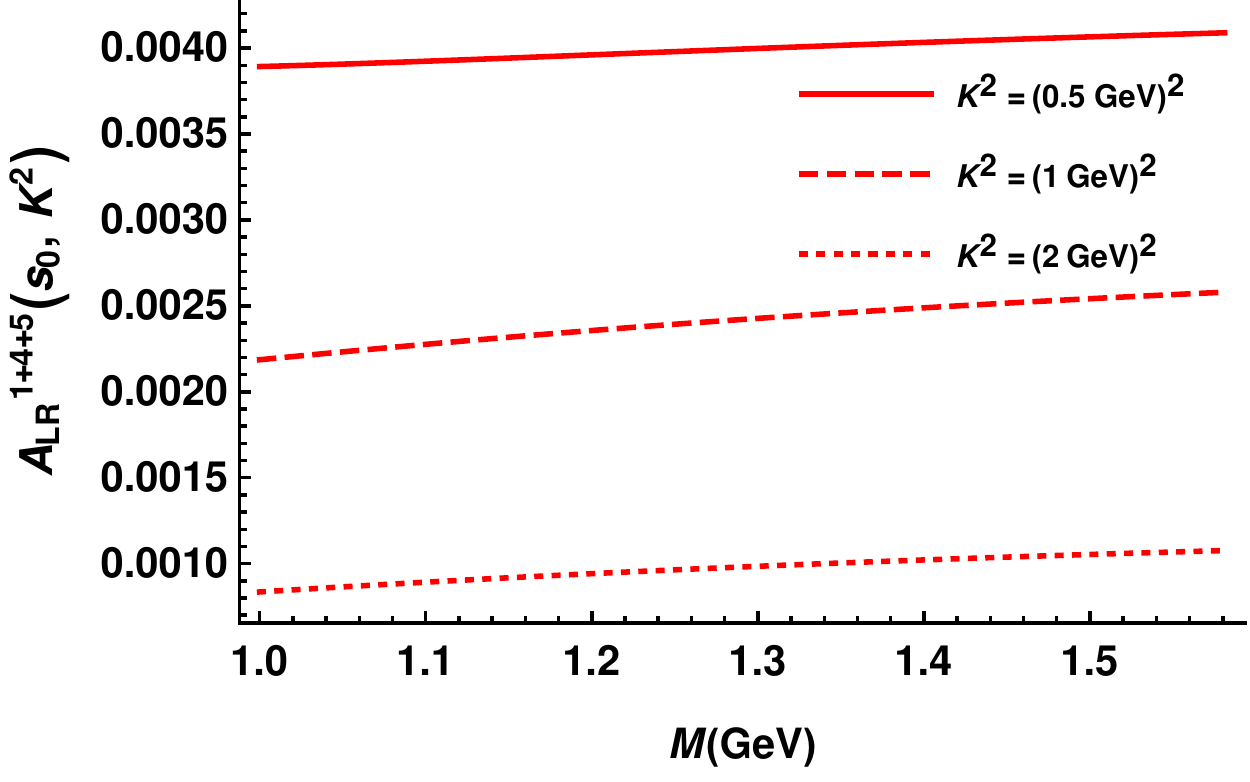}
    \caption{}
   \end{subfigure}
    \caption{The physical FF, $A_{LR}(s_0,K^2)$ is calculated from the combination of $F_{LR}^{1}$, $F_{LR}^{4}$ and $F_{LR}^{5}$ employing proton DAs. Left panel: $A_{LR}^{1+4+5}(s_0,K^2)$ vs $K^2$ is shown for three values of $s_0= (1.4 \text{ GeV})^2$(violate dotted), $s_0= (1.44 \text{ GeV})^2$(red solid) and $s_0= (1.5 \text{ GeV})^2$ (blue dashed) at the Borel Mass, $M^2= 2 \text{ GeV}^2$. Right Panel: $A_{LR}^{1+4+5}(s_0,K^2)$ vs $M$ is shown for three values of $K^2= 0.5 \text{ GeV}^2$(red solid), $K^2= 1 \text{ GeV}^2$(red dashed) and $K^2= 2 \text{ GeV}^2$ (red dotted) at the continuum threshold, $s_0=(1.44 \text{ GeV})^2$.}
    \label{LRp}
\end{figure}

The physical FFs, $A_{\Gamma\Gamma'}$ are studied as a function of $K^2=-k^2$ and the Borel mass, $M$ at $P'^2=m_e^2=0$. Using Eq.(\ref{fphys}) and Eq.(\ref{relations2}), one can see that the physical form factors are proportional to $W_2(K^2)$ which can be calculated using other combinations of $F_{\Gamma\Gamma'}^n$ as well. We have found that the most stable one against the Borel mass is obtained from the the combination of $F_{\Gamma\Gamma'}^1$, $F_{\Gamma\Gamma'}^4$, and $F_{\Gamma\Gamma'}^5$ as defined in Eq.(\ref{fphys}). We thus choose to show this explicitly in Fig.(\ref{LLp}) and Fig(\ref{LRp}). We would also like to remark that a direct comparison of the form factors obtained here with those obtained when the proton is interpolated and photon DAs are used is not possible. The simple reason being that in the present case, the photon is far off-shell while in the previous case photon is on-shell and hence, in our view, the form factors so obtained in the previous case are better suited for a phenomenological analysis.
In the present case, some kind of judicious extrapolation would be required. There is another issue that is worth pointing out. When employing proton (or nucleon) DAs while computing the electromagnetic form factors of the nucleons, it has been observed that the choice of the interpolation current plays a crucial role \cite{Braun:2006hz}. For some choice(s), particular form factors simply don't actually show up in the correlator calculation. In the case at hand, the four quark operator, with the positron field factored out, can be thought of as an analog of an interpolating current. Thus, it seems that differences or ambiguities similar to the above discussion are perhaps at play even here as the form factor $A_{LL}$ in Fig.(\ref{LLp}) is about an order of magnitude smaller than $A_{LR}$, and also with the form factors determined with photon DAs.

\section{Discussion and Conclusions}
\label{discussion}
In this work, we have computed the form factors involved in the proton decay to a positron and a photon using the LCSR framework. This should be viewed as a complimentary approach to lattice calculations, though, to the best of our knowledge, no lattice study exists for proton to gamma transition. This decay mode has not attracted much attention. However, as briefly discussed in \cite{Silverman:1980ha}, the branching ratio for this mode is expected to be smaller than the $p\to \pi e^+$ mode by a factor $\mathcal{O}(1/(\text{few tens}))$. This is not a huge suppression and keeping in mind that the nuclear absorption effects are not going to affect the radiative mode, it is important to remain optimistic about this mode. The next important task is to have the relevant form factors computed in a reliable fashion. Choosing to work in the framework of light cone sum rules, these form factors can be calculated either by interpolating the proton state and using the photon DAs or by interpolating the photon state and using the proton DAs. We have considered both these scenarios one by one. The physical form factors that would enter the decay rate for the radiative process can be determined from different combinations of hadronic functions that can be systematically computed. In the case when photon DAs are emplyed for computing the correlation functions, we find that the condensate contributions do turn out to be important and dominant for specific hadronic functions. Thus, not having considered these would have led to erroneous results. In the case of proton DAs, at the order in twist employed for the present calculations, condensate contributions do not appear. For both the cases, we have explicitly shown the form factors for the combinations that present the best Borel stability. As we have briefly discussed above, in our opinion, the form factors determined using photon distribution amplitudes (Case-1) are more trustworthy. This also motivates for more detailed studies employing proton DAs in order to gain better insight into the issues, including investigating the effect of the condensates at twist-4 and higher. In the first case i.e., when photon DAs are employed, the calculations performed do not include three particle twist-3 contributions. This is justified at the level of precision needed at present as these contributions are expected to be about an order of magnitude smaller than those already included since two-particle twist-3 contributions are found to be typically an order of magnitude smaller. \\
The detailed expressions for all the hadronic correlators are listed in the appendices and are exact in the sense that they are written for non-zero positron mass and without assuming $k^2=0$. While computing the amplitude we have assumed positron to be massless. Some extra contributions will arise due to non-zero lepton mass while manupulating Eq.(3) and Eq.(7). Thus, with very little effort, these can be utilised to compute form factors and thus branching ratio if there is $\mu^+$ instead of $e^+$ in the final state. Some  Final states with second generation particles may be favoured channels in scenarios where the scalar mediated contribution dominates over the gauge mediated one (see for example \cite{Patel:2022wya} for a recent study pointing out this feature). The radiative modes thus become equally important and can provide complimentary information about the details of the underlying high energy theory.

\begin{appendix}
\numberwithin{equation}{section}
\section{Distribution Amplitudes (DAs)}
\label{DAs}
\subsection{Proton DAs} Considering the Lorentz covariance, parity and spin of the nucleon, the matrix element of three quark operator between the vacuum and the nucleon state can be decomposed into 24 invariant functions in general. These functions are related to the light cone distribution amplitudes of the proton (see \cite{Braun:2000kw} for the details). At twist-3, there are three DAs (Eq.(\ref{pda})): the vector, $V_1$, the axial-vector, $A_1$  and the tensor, $T_1$. The explicit conformal expansion of these DAs are:
\begin{equation}
    V_1(\alpha_i,\mu) = 120 \alpha_1\alpha_2\alpha_3\left[\phi_3^0(\mu)+\phi_3^+(\mu)(1-3\alpha_3)\right]
\end{equation}
\begin{equation}
    A_1(\alpha_i,\mu) = 120\alpha_1\alpha_2\alpha_3(\alpha_2-\alpha_1)\phi_3^-(\mu)
\end{equation}
\begin{equation}
    T_1(\alpha_i,\mu)= 120 \alpha_1\alpha_2\alpha_3\left[\phi_3^0(\mu)+\frac{1}{2}(\phi_3^--\phi_3^+)(\mu)(1-\alpha_3)\right]
\end{equation}
Here, $\alpha_i$ ($i=1,2,3$) are the momentum fractions of the nucleon momentum carried by the three quarks. $\phi_3^0(\mu)$, $\phi_3^+(\mu)$, and $\phi_3^-(\mu)$ are the renormalisation scale, $\mu$, dependent coefficients. They are available from QCD sum rules and are provided in Appendix-\ref{appendixB}
 \subsection{Photon DAs}
The photon DAs are defined as the vacuum expectation value of the non-local quark-antiquark plus $n$ gluons operator (when $n\geq 0$) with light-like separations. We have considered only the two particle i.e. quark-antiquark DAs of twist-2 and twist-3 in the present work which are defined as follows: 
 \begin{enumerate}
     \item \underline {Twist-2 DAs:} At twist-2, we have only one two-particle DA, $\phi_\gamma(u)$ which is defined as
     \begin{equation}
           \left<\gamma(k)\left|\bar q(0) \sigma_{\rho\sigma} q(x)\right|0\right> = -i e_q \left<\bar q q\right> (\epsilon_\rho k_\sigma - \epsilon_\sigma k_\rho) \int_0^1 du e^{i \bar u k.x} \chi \phi_\gamma(u).
        \end{equation}
        Here, $\left<\bar q q\right>$ is the quark condensate, $\epsilon_\mu$ is the polarisation vector of the photon, $e_q = Q_q e$ is the electric charge of the quark and $\chi$ is the magnetic susceptibility. $u$ and $\bar u = 1-u$ are the momentum fractions carried by the quark and anti-quark, respectively. $\phi_\gamma(u)$ is the photon DA of twist-2. The asymptotic form of this DA is
        \begin{equation}
            \phi_\gamma^{asy}(u)=6u(1-u)
        \end{equation}
     \item \underline {Twist-3 DAs:} At twist-3, there are four DAs out of which two are two-particle DAs and two are three particle DAs. The two particle DAs are defined as
          \begin{equation}
          \left<\gamma(k)\left|\bar q(0) \gamma_\mu q(x)\right|0\right> = e_q f_{3\gamma}\left(\epsilon^*_\mu -k_\mu \frac{\epsilon^*x}{kx}\right) \int_0^1 du e^{i \bar u k.x}\psi^{v}(u,\mu)
        \end{equation}
        \begin{equation}
        \left<\gamma(k)\left|\bar q(0) \gamma_\mu \gamma_5 q(x)\right|0\right> = \frac{1}{4}e_q f_{3\gamma}\epsilon_{\mu\nu\alpha\beta}k^\alpha x^\beta \epsilon^{*\mu}  \int_0^1 du e^{i \bar u k.x}\psi^{a}(u,\mu)
        \end{equation}
        where, $f_{3\gamma}$ provides a natural mass scale for twist-3 DAs $\psi^v(u)$ and $\psi^a(u)$. The explicit form of these DAs are:
        \begin{equation}
            \psi^{(v)}(u)=5\left(3\xi^2-1\right)+\frac{3}{64}\left(15 \omega_\gamma^V-5\omega_\gamma^A\right)\left(3-30\xi^2+35\xi^4\right)
        \end{equation}
        \begin{equation}
            \psi^{(a)}(u)=\left(1-\xi^2\right)\left(5\xi^2-1\right)\frac{5}{2}\left(1+\frac{9}{16} \omega_\gamma^V-\frac{3}{16}\omega_\gamma^A\right)
        \end{equation}
    where,  $\xi=2u-1$ and $\omega_\gamma^V$ $\&$ $\omega_\gamma^A$ corresponds to the local operators of dimension six. The values of these constants are provided in Appendix-\ref{appendixB}.
     Twice the integral of $\psi^v(\alpha)$ over $\alpha$ from 0 to $u$ is defined as $\bar \psi^v(u)$ and is given by
 \begin{align}
     \bar \psi^v(u) &= 2\int_0^u d\alpha \psi^v(\alpha) \nonumber \\ &= -20 u \bar u \xi + \frac{15}{16} \left(\omega_\gamma^A-3\omega_\gamma^V\right)u \bar u \xi \left(7 \xi^2-3\right)
 \end{align}
For photon DAs of higher twist and DAs corresponding to three or more particles, one can look at \cite{Ball:2002ps}.
 \end{enumerate}

\section{Correlation functions for Case-1 (employing photon DAs)}
\label{cf}
In this appendix, we collect the analytic results of the correlation functions $\Pi_{\Gamma\Gamma'}^{r}(p_e,p_p)$ computed in QCD. 
\begin{align}
    \Pi_{LL}^{QCD,T}(p_e,p_p)= -e\left<\bar q q\right>\int_0^1 du &\left[\frac{3Q_u \chi}{16 \pi^2}\phi_\gamma(u) P^2 \text{ln}(-P^2)+\frac{f_{3\gamma}(Q_u-Q_d)}{6}\left\{\frac{1}{P^2}\left(1+\frac{m_0^2}{4P^2}\right)\left(u \psi^{(v)}(u)-\frac{\bar \psi^{(v)}(u)}{2}\right) \right.\right. \nonumber \\ &\left.\left.+\frac{\psi^{a}(u)}{2P^4}\left(1+\frac{m_0^2}{2P^2}\right)\left(u k.p_p-p_p^2\right)\right\}\right]
\end{align}
\begin{align}
    \Pi_{LL}^{QCD,TP}(p_e,p_p)= \frac{e\left<\bar q q\right>}{6}(Q_u-Q_d)\int_0^1 du &\left[\left<\bar q q\right> \chi \frac{\phi_\gamma(u)}{P^2}\left(1+\frac{m_0^2}{4P^2}\right)-f_{3\gamma}\frac{\bar \psi^v(u)}{P^4}\left(1+\frac{m_0^2}{2P^2}\right)\right]
\end{align}
\begin{align}
    \Pi_{LL}^{QCD,KK}(p_e,p_p)= -\frac{e\left<\bar q q\right>^2 \chi}{6}(Q_u-Q_d)\int_0^1 du  \frac{\phi_\gamma(u)}{P^2}\left(1+\frac{m_0^2}{4P^2}\right)
\end{align}
\begin{align}
    \Pi_{LL}^{QCD,V}(p_e,p_p)= \frac{e\left<\bar q q\right>}{6}(Q_u-Q_d)\int_0^1 du  &\left[\left<\bar q q\right> \chi \frac{\phi_\gamma(u)}{P^2}\left(1+\frac{m_0^2}{4P^2}\right)- f_{3\gamma} \frac{\bar \psi^v(u)}{P^4}\left(1+\frac{m_0^2}{2P^2}\right)\right]uk^2
\end{align}
\begin{align}
    \Pi_{LL}^{QCD,VP}(p_e,p_p)= -\frac{ef_{3\gamma}\left<\bar q q\right>}{6}(Q_u-Q_d)\int_0^1 du  &\left[ \frac{\psi^v(u)}{P^2}\left(1+\frac{m_0^2}{4P^2}\right)- (k.p_p-uk^2) \frac{ \psi^a(u)}{2P^4}\left(1+\frac{m_0^2}{2P^2}\right)\right]
\end{align}
\begin{align}
    \Pi_{LL}^{QCD,K}(p_e,p_p)= \frac{ef_{3\gamma}\left<\bar q q\right>}{6}(Q_u-Q_d)\int_0^1 du  &\left[ \frac{1}{P^2}\left(1+\frac{m_0^2}{4P^2}\left(u\psi^v(u)+\frac{\bar\psi^v(u)}{2}\right)\right)\nonumber\right.\\ &+ \left.\left( 2u\bar\psi^v(u)-u(p_p.k)\psi^a(u)\right) \frac{1}{P^4}\left(1+\frac{m_0^2}{2P^2}\right)\right]
\end{align}
\begin{align}
    \Pi_{LL}^{QCD,KP}(p_e,p_p)= -\frac{ef_{3\gamma}\left<\bar q q\right>}{6}(Q_u-Q_d)\int_0^1 du   \frac{\bar \psi^v(u)}{P^4}\left(1+\frac{m_0^2}{2P^2}\right)
\end{align}
\begin{align}
    \Pi_{LL}^{QCD,KPP}(p_e,p_p)= -\frac{ef_{3\gamma}\left<\bar q q\right>}{12}(Q_u-Q_d)\int_0^1 du   \frac{ \psi^a(u)}{P^4}\left(1+\frac{m_0^2}{2P^2}\right)
\end{align}
\begin{align}
    \Pi_{LL}^{QCD,P}(p_e,p_p)= \frac{ef_{3\gamma}\left<\bar q q\right>}{12}(Q_u-Q_d)\int_0^1 du   \frac{ uk^2\psi^a(u)}{P^4}\left(1+\frac{m_0^2}{2P^2}\right)
\end{align}
\begin{align}
    \Pi_{LL}^{QCD,KKP}(p_e,p_p)= \frac{ef_{3\gamma}\left<\bar q q\right>}{12}(Q_u-Q_d)\int_0^1 du   \frac{ u\psi^a(u)}{P^4}\left(1+\frac{m_0^2}{2P^2}\right)
\end{align}
\begin{align}
    \Pi_{LR}^{QCD,T}(p_e,p_p)= \frac{e\left<\bar q q\right>}{6}\int_0^1 du  &\left[\frac{Q_d}{8\pi^2}\chi \phi_\gamma(u)\left(5P^2+2u(p_p.k-uk^2)\right)\text{ln}(-P^2) + f_{3\gamma} Q_u (p_p^2-up_p.k)\frac{\psi^a(u)}{P^4}\left(1+\frac{m_0^2}{2P^2}\right)\right]
\end{align}
\begin{align}
    \Pi_{LR}^{QCD,KPP}(p_e,p_p)= -\frac{e\left<\bar q q\right>}{6}\int_0^1 du  &\left[\frac{Q_d}{4\pi^2}\chi \phi_\gamma(u)\text{ln}(-P^2) + f_{3\gamma} Q_u \frac{\psi^a(u)}{P^4}\left(1+\frac{m_0^2}{2P^2}\right)\right]
\end{align}
\begin{align}
    \Pi_{LR}^{QCD,KKP}(p_e,p_p)= \frac{e\left<\bar q q\right>}{6}\int_0^1 du  &\left[\frac{Q_d}{4\pi^2}\chi u\phi_\gamma(u)\text{ln}(-P^2) + f_{3\gamma} Q_u \frac{u\psi^a(u)}{P^4}\left(1+\frac{m_0^2}{2P^2}\right)\right]
\end{align}
\begin{align}
    \Pi_{LR}^{QCD,P}(p_e,p_p)= \frac{e\left<\bar q q\right>}{3}\int_0^1 du  &\left[\frac{Q_d}{8\pi^2}\chi uk^2\phi_\gamma(u)\text{ln}(-P^2) - f_{3\gamma} Q_u\left\{\frac{\psi^v(u)}{P^2}\left(1+\frac{m_0^2}{4P^2}\right) \nonumber \right.\right. \\ &+\left.\left.\left(\bar \psi^v(u)(k.p_p-uk^2)-\frac{uk^2 \psi^a(u)}{2}\right) \frac{1}{P^4}\left(1+\frac{m_0^2}{2P^2}\right)\right\}\right]
\end{align}
\begin{align}
    \Pi_{LR}^{QCD,K}(p_e,p_p)= -\frac{e\left<\bar q q\right>}{3}\int_0^1 du  &\left[\frac{Q_d}{8\pi^2}\chi u(p_p.k)\phi_\gamma(u)\text{ln}(-P^2) - f_{3\gamma} Q_u\left\{\left(u\psi^v(u)+\frac{\bar\psi^v(u)}{2}\right)\frac{1}{P^2}\left(1+\frac{m_0^2}{4P^2}\right) \nonumber \right.\right. \\ &+\left.\left.\left(u(k.p_p-uk^2)\bar \psi^v(u)-\frac{u(p_p.k) \psi^a(u)}{2}\right) \frac{1}{P^4}\left(1+\frac{m_0^2}{2P^2}\right)\right\}\right]
\end{align}
\begin{align}
    \Pi_{LR}^{QCD,VP}(p_e,p_p)= \frac{e\left<\bar q q\right>}{6}\int_0^1 du  &\left(p_p.k-uk^2\right)\left[\frac{Q_d}{4\pi^2}\chi \phi_\gamma(u)\text{ln}(-P^2) + f_{3\gamma} Q_u \frac{\psi^a(u)}{P^4}\left(1+\frac{m_0^2}{2P^2}\right)\right]
\end{align}
\begin{align}
    \Pi_{LR}^{QCD,TP}(p_e,p_p)= e\int_0^1 du  &\left[\frac{\left<\bar q q\right>^2 \chi Q_u}{3} \frac{\phi_\gamma(u)}{P^2}\left(1+\frac{m_0^2}{4P^2}\right) + \frac{f_{3\gamma} Q_u}{16\pi^2}\psi^a(u)\text{ln}(-P^2) \right]
\end{align}
\begin{align}
    \Pi_{LR}^{QCD,V}(p_e,p_p)= e\int_0^1 du  &\left[\frac{\left<\bar q q\right>^2 \chi Q_u}{3}\frac{(p_p.k)\phi_\gamma(u)}{P^2}\left(1+\frac{m_0^2}{4P^2}\right) + \frac{f_{3\gamma}}{16\pi^2}\left\{\frac{1}{3}\left((7Q_u+Q_d)\psi^v(u)P^2  \nonumber \right.\right.\right. \\ &-\left.\left.\left. (Q_u+Q_d)\bar\psi^v(u)(p_p.k-uk^2)\right)+Q_u \psi^a(u)(p_p.k)\right\}\text{ln}(-P^2)\right]
\end{align}
\begin{align}
    \Pi_{LR}^{QCD,PK}(p_e,p_p)= -e\int_0^1 du  &\left[\frac{\left<\bar q q\right>^2 \chi Q_u}{3}\frac{\phi_\gamma(u)}{P^2}\left(1+\frac{m_0^2}{4P^2}\right) + \frac{f_{3\gamma}}{16\pi^2}\left\{\left\{\frac{1}{3}\left(2(Q_u+Q_d)u\psi^v(u)  \nonumber\right. \right.\right.\right. \\ &+\left.\left.\left.\left. (7Q_u+Q_d)\bar\psi^v(u)\right)+Q_u \psi^a(u)\right\}\text{ln}(-P^2)+\frac{2(Q_u+Q_d)}{3P^2}u\left(p_p.k-uk^2\right)\bar\psi^v(u)\right\}\right]
\end{align}
\begin{align}
    \Pi_{LR}^{QCD,PP}(p_e,p_p)= \frac{ef_{3\gamma}}{24\pi^2}(Q_u+Q_d)\int_0^1 du  &\left[\psi^v(u)\text{ln}(-P^2)+ (p_p.k-uk^2)\frac{\bar\psi^v(u)}{P^2} \right]
\end{align}
\begin{align}
    \Pi_{LR}^{QCD,KP}(p_e,p_p)= -\frac{ef_{3\gamma}}{24\pi^2}(Q_u+Q_d)\int_0^1 du  &\left[\left(u\psi^v(u)+\bar\psi^v(u)\right)\text{ln}(-P^2)+ (p_p.k-uk^2)\frac{u\bar\psi^v(u)}{P^2} \right]
\end{align}
\begin{align}
    \Pi_{LR}^{QCD,KK}(p_e,p_p)= \frac{ef_{3\gamma}}{24\pi^2}\int_0^1 du  &u^2\left[\left\{(Q_u+Q_d)\psi^v(u)+(4Q_u+Q_d)\frac{\bar\psi^v(u)}{u}\right\}\text{ln}(-P^2)+ (Q_u+Q_d)(p_p.k-uk^2)\frac{\bar\psi^v(u)}{P^2} \right]
\end{align}
Here, $P^2=(p_p-uk)^2= (p_e+uk)^2 = \bar u p_p^2-u P_e^2 - u\bar u k^2 $. The remaining correlation functions does not appear in QCD calculations upto the twist accuary we have considered.  We perform the Borel transform on $p_p^2$ to get the final sum rules.

\section{Correlation functions for Case-2 (employing proton DAs)}
\label{cfpda}
In this appendix, we collect the analytic results for the correlation functions $F_{\Gamma\Gamma'}^n (p_p,k)$ computed in QCD. 
\begin{equation}
    F_{LL}^{3,QCD}(p_p,k)= -\frac{e m_p^2}{2} \int \mathcal{D}\alpha_i T_1(\alpha_i) \left[\frac{\alpha_3 Q_d}{(k-\alpha_3 p_p)^2}+\frac{\alpha_1 Q_u}{(k-\alpha_1 p_p)^2}\right]
\end{equation}
\begin{equation}
    F_{LL}^{4,QCD}(p_p,k)= -\frac{e m_p^2}{2} \int \mathcal{D}\alpha_i T_1(\alpha_i) \left[\frac{ Q_d}{(k-\alpha_3 p_p)^2}+\frac{ Q_u}{(k-\alpha_1 p_p)^2}\right]
\end{equation}
\begin{equation}
    F_{LL}^{5,QCD}(p_p,k)= \frac{e m_p^2}{2} \int \mathcal{D}\alpha_i T_1(\alpha_i) \left[\frac{\alpha_1 Q_u}{(k-\alpha_1 p_p)^2}-\frac{2 \alpha_3 Q_d}{(k-\alpha_3 p_p)^2}\right]
\end{equation}
\begin{equation}
    F_{LL}^{6,QCD}(p_p,k)= \frac{3Q_de m_p^2}{2} \int \mathcal{D}\alpha_i \frac{T_1(\alpha_i)}{(k-\alpha_3 p_p)^2}
\end{equation}

\begin{equation}
    F_{LR}^{1,QCD}(p_p,k)= \frac{e m_p^2}{2} \int \mathcal{D}\alpha_i\left[ \frac{\left(V_1(\alpha_i)+A_1(\alpha_i)\right)Q_d}{(k-\alpha_3 p_p)^2}-\frac{\left(V_1(\alpha_i)-A_1(\alpha_i)\right)Q_u}{(k-\alpha_1 p_p)^2}\right]
\end{equation}
\begin{align}
    F_{LR}^{3,QCD}(p_p,k)= -\frac{e}{2} \int \mathcal{D}\alpha_i&\left[ \frac{\left(V_1(\alpha_i)+A_1(\alpha_i)\right)Q_d\left(2p_p.k-\alpha_3 m_p^2\right)}{2(k-\alpha_3 p_p)^2} \right. \nonumber \\ &+ \left.\frac{\left(V_1(\alpha_i)-A_1(\alpha_i)\right)Q_u\left(2\alpha_1m_p^2-p_p.k\right)}{(k-\alpha_1 p_p)^2}\right]
\end{align}
\begin{equation}
    F_{LR}^{4,QCD}(p_p,k)= -\frac{em_p^2}{2} \int \mathcal{D}\alpha_i\left[ \frac{\left(V_1(\alpha_i)+A_1(\alpha_i)\right)Q_d}{2(k-\alpha_3 p_p)^2}+\frac{\left(V_1(\alpha_i)-A_1(\alpha_i)\right)Q_u}{(k-\alpha_1 p_p)^2}\right]
\end{equation}
\begin{equation}
    F_{LR}^{6,QCD}(p_p,k)= -\frac{em_p^2}{2} \int \mathcal{D}\alpha_i\left[ \frac{\left(V_1(\alpha_i)+A_1(\alpha_i)\right)Q_d}{2(k-\alpha_3 p_p)^2}-\frac{\left(V_1(\alpha_i)-A_1(\alpha_i)\right)Q_u}{(k-\alpha_1 p_p)^2}\right]
\end{equation}
The remaining correlation functions does not appear in QCD calculations upto the twist accuary we have considered. 
In this case, the Borel transformation will be performed on $P'^2=(p_p-k)^2=p_e^2$.

\section{Conventions, Definitions and Identities}
\label{appendixA}
\subsection{Definitions and Conventions}
As discussed in Section-\ref{case1}, the interpolation current for proton state is not unique. The Ioffe current, $\chi(x)$ as defined in Eqn.\ref{ioffe} is the linear combination of $\chi_1(x)$ and $\chi_2(x)$ defined in Eqn-\ref{chi12} as,
\begin{equation}
    \chi(x)=2(\chi_2-\chi_1)
\end{equation}
such that, \begin{equation}
\label{chi}
    \left<0\left|\chi(0)\right|p(p_p)\right> = m_p \lambda_p u_p(p_p).
\end{equation}
There is another interpolation current as a linear combination of these two currents defined as,
\begin{align}
    \chi'(x)&= 2(\chi_2+\chi_1) \nonumber \\
    &= \frac{1}{2}\epsilon^{abc}\left(u^{Ta}(x)C\sigma_{\mu\nu}u^b(x)\right)\sigma^{\mu\nu}\gamma_5d^c(x)
\end{align}
such that,
 \begin{equation}
 \label{chip}
    \left<0\left|\chi'(0)\right|p(p_p)\right> = m_p \lambda_p' u_p(p_p)
\end{equation}

\subsection{Useful identities and integrals}
\begin{itemize}
    \item \textbf{Identities:}
    \begin{enumerate}
        \item For $\sigma = \frac{i}{2}\left[\gamma^\rho,\gamma^\sigma\right]$, 
        \begin{equation}
            \gamma^\alpha \sigma^{\rho\sigma}=2 i g^{\alpha\rho}\gamma^\sigma - 2i \gamma^\rho g^{\alpha\sigma}+\sigma^{\rho\sigma}\gamma^\alpha
        \end{equation}
 
    \item Chisholm Identity:
    \begin{equation}
        \gamma^\alpha\gamma^\beta \gamma^\mu = g^{\alpha\beta}\gamma^\mu - g^{\alpha\mu}\gamma^\beta + g^{\beta \mu}\gamma^\alpha-i \epsilon^{\alpha\beta\mu\nu}\gamma_\nu \gamma_5
    \end{equation}
    \end{enumerate}
   
\item \textbf{Integrals:} In D dimensions using dimensional regularisation, the formula for general integrations involved in the correlation function is given by \cite{Khodjamirian:2020btr},
\begin{equation}
    \int d^Dx e^{ipx}\frac{1}{\left(x^2\right)^n}=\left(-i\right)\left(-1\right)^n 2^{(D-2n)}\pi^{D/2}\left(-p^2\right)^{n-D/2}\frac{\Gamma\left(D/2-n\right)}{\Gamma(n)}
\end{equation}
for $n\geq1$ ,$p^2<0$. On differentiating it over the four-momentum $p_\alpha$, we get the desired form of the integrals involved in our calculations.
\begin{align}
    &\int d^4x e^{ipx}\frac{x_\alpha}{x^4} = 2\pi^2 \frac{p_\alpha}{p^2}, \hspace{4 cm}  \int d^4x e^{ipx}\frac{x_\alpha}{x^2} = 8\pi^2 \frac{p_\alpha}{p^4} \nonumber \\ 
    &\int d^4x e^{ipx}\frac{x_\alpha x_\beta}{x^4} = -\frac{2i\pi^2}{p^2}\left(g_{\alpha\beta}-\frac{2p_\alpha p_\beta}{p^2}\right), \hspace{1cm}  \int d^4x e^{ipx}\frac{x_\alpha x_\beta}{x^2} = -\frac{8i\pi^2}{p^4}\left(g_{\alpha\beta}-\frac{4p_\alpha p_\beta}{p^2}\right) \nonumber \\ &\int d^4x e^{ipx}\frac{x_\alpha}{x^6} =\frac{-\pi^2}{4} p_\alpha ln(-p^2), \hspace{2.5cm}   \int d^4x e^{ipx}\frac{1}{x^6} =\frac{-i\pi^2}{8} p^2 ln(-p^2) \nonumber \\ &  \int d^4x e^{ipx}\frac{x_\alpha x_\beta}{x^8} =\frac{-i\pi^2}{48} \left(p^2g_{\alpha\beta}+2p_\alpha p_\beta\right) ln(-p^2) \nonumber \\ &  \int d^4x e^{ipx}\frac{x_\alpha x_\beta x_\mu}{x^8} = \frac{\pi^2}{24}\left(\frac{2p_\alpha p_\beta p_\mu}{p^2}-\left(p_\alpha g_{\beta\mu}+p_\beta g_{\alpha\mu}+p_\mu g_{\alpha\beta}\right)ln(-p^2)\right)
\end{align}
Here, the divergent terms which are proportional to $p^2$ are omitted as they goes to zero after Borel transformaion.
\end{itemize}

\subsection{Borel Transformations}
As listed in Appendix-\ref{cf} and Appendix-\ref{cfpda}, the correlation functions calculated in QCD involves,
\begin{equation}
    P^2= (p_p-uk)^2= (p_e+uk)^2 = \bar u p_p^2-u P_e^2 - u \bar u k^2
\end{equation}
with $P_e^2=-p_e^2$ and $\bar u=1-u$ in case-1 and
\begin{equation}
    (k-\alpha p_p)^2= \alpha P'^2 - \bar \alpha K^2 - \alpha \bar \alpha m_p^2 
\end{equation}
with $\alpha=\{\alpha_1,\alpha_3\}$, $K^2=-k^2$ and $P'^2=(p_p-k)^2$ in case-2.
To calculate the final sum rules, one need to find the imaginary part of the correlation functions collected in Appendix-\ref{cf} and Appendix-\ref{cfpda} and substitute them in Eq.(27) and Eq.(44), which are obtained by performing the Borel transformations on the momentum trasferred square i.e. $p_p^2$ and $P'^2=(p_p-k)^2$ for case-1 and case-2, respectively. To incorporate that, one need to make the following substitutions in the correlation functions of case-1,
\begin{equation}
    \int_0^1 du \frac{F(u)}{P^2} G(u,s) \rightarrow -\int_0^{u_0} du \frac{F(u)}{\bar u}e^{\frac{-\tilde s}{M^2}} G(u,\tilde s) 
\end{equation}
\begin{equation}
    \int_0^1 du \frac{F(u)}{P^4} G(u,s) \rightarrow \frac{e^{\frac{-s_0}{M^2}}F(u_0)G(s_0,u_0)}{P_e^2}+\int_0^{u_0} du \frac{F(u)}{\bar u^2}\frac{e^{\frac{-\tilde s}{M^2}}}{M^2} \left(G(u,\tilde s)-M^2 \frac{\partial}{\partial \tilde s}G(u,\tilde s)\right) 
\end{equation}
\begin{align}
    \int_0^1 du \frac{F(u)}{P^6} G(u,s)& \rightarrow -\int_0^1 du \frac{F(u)}{2 \bar u^2}\left[e^{\frac{-s_0}{M^2}}G(u,s_0)\frac{\partial}{\partial s_0}\left( \delta(\bar u s_0-u P_e^2)\right)\right]\nonumber\\ &+\int_0^1 \frac{F(u)}{2 \bar u^2}\left[\frac{\partial}{\partial s}\left(e^{\frac{-s}{M^2}}G(u,s)\right)\delta(\bar u s-uP_e^2)\right] \nonumber \\ &- \int_0^{u_0} du \frac{F(u)}{2\bar u^3}\frac{\partial^2}{\partial \tilde s^2}\left(e^{\frac{-\tilde s}{M^2}}G(u,\tilde s)\right)
\end{align}
with
\begin{equation}
    \tilde s=\frac{u P_e^2}{\bar u} \hspace{1.5cm} \text{and}\hspace{1.5cm} u_0 = \frac{s_0}{s_0+P_e^2}.
\end{equation}
In these substitutions we put $s=p_p^2$ and $k^2=0$ as the photon is onshell. 
These substitutions are consistent with \cite{Braun:2006hz}.\\
For case two, the subsbtitution reads as,
\begin{equation}
    \int \mathcal{D}\alpha_i \frac{F(\alpha_i)}{(k-\alpha p_p)^2}\rightarrow -\int_{\alpha_0}^1 \mathcal{D}\alpha_i \frac{F(\alpha_i)}{\alpha}e^{\frac{-s_1}{M^2}}
\end{equation}
with $\alpha=\{\alpha_1,\alpha_3\}$, $\mathcal{D}\alpha_i = d\alpha_1d\alpha_2 d\alpha_3\delta(1-\alpha_1-\alpha_2-\alpha_3)$, 
\begin{equation}
    s_1 = \frac{\bar \alpha K^2 + \alpha \bar \alpha m_p^2}{\alpha}
\end{equation}
and 
\begin{equation}
    \alpha_0 = -\frac{K^2-m_p^2+s_0}{2m_p^2}+\frac{\sqrt{(K^2+s_0)^2+m_p^4-2m_p^2(s_0-K^2)}}{2m_p^2}.
\end{equation}
Here, $s=(p_p-k)^2$ and $K^2=-k^2$.
 \section{Values of parameters used}
\label{appendixB}
In this appendix, we collect all the numerical values of the parameters used for both case-1 and case-2 during numerical analysis.
\begin{center}
    \begin{tabular}{|c||c|c|c|}
    \hline
 \textbf{S.No.}&  \textbf{ Parameter} & \textbf{Value Used} & \textbf{Reference}\\
 \hline\hline
  \textbf{1. }& Proton mass ($m_p$)& 0.938 GeV &\cite{Zyla:2020zbs}\\
    \hline
     \textbf{2.} &Fine Structure Constant $\left(\alpha=\frac{e^2}{4\pi}\right)$ & $\frac{1}{137}$ & \cite{Zyla:2020zbs} \\
    \hline
     \textbf{  3.} &Quark condensate ($\left<\bar q q\right>$) & $-((256\pm2)\text{MeV})^3$ &\cite{Haisch:2021hvj}\\
    \hline
     \textbf{  4.} & $m_0^2$ & $(0.8)\pm0.2 \text{GeV}^2$  & \cite{Haisch:2021hvj}\\
  
    \hline
     \textbf{5. } & Magnetic Susceptibility ($\chi$)&$( 3.08\pm 0.02) \text{GeV}^2$  & \cite{Ball:2002ps}\\
    \hline
       \textbf{6. } & $f_{3\gamma}$ & $-(4\pm2).10^{-3}\text{GeV}^2$  &\cite{Ball:2002ps}\\
    \hline
       \textbf{7.} & $\omega_\gamma^v$& $3.8\pm1.8$  &\cite{Ball:2002ps}\\
    \hline
      \textbf{8. } &$\omega^a_\gamma$ & $-2.1\pm 1.0$ &\cite{Ball:2002ps}\\
    \hline
       \textbf{ 9. }& $\lambda_p$& $(5.4\pm1.9).10^{-2} \text{GeV}^2$  &\cite{Braun:2006hz}\\
       \hline
    \textbf{10. } & $\lambda_p'$ & $-(2.7\pm0.9).10^{-2}\text{GeV}^2$ &\cite{Braun:2006hz}\\
    \hline
     \textbf{11. } & $\phi_3^0(\mu=1\text{GeV})$ & $(5.3\pm0.5).10^{-3} \text{GeV}^2$&\cite{Braun:2000kw}\\
    \hline
      \textbf{12. } & $\tilde \phi_3^+(\mu=1\text{GeV}) = \frac{\phi_3^+}{\phi_3^0}$ &$1.1\pm0.3$ &\cite{Braun:2000kw}\\
    \hline
      \textbf{13. } & $\tilde \phi_3^-(\mu=1\text{GeV}) = \frac{\phi_3^-}{\phi_3^0}$ &$4.0\pm1.5$ &\cite{Braun:2000kw}\\
    \hline
\end{tabular}
\end{center}

\end{appendix}
\bibliography{Proton_decay}{}
\bibliographystyle{unsrt}
\end{document}